\documentclass[a4paper,fleqn,usenatbib,useAMS]{mnras}

\usepackage{graphicx}	
\usepackage{amsmath}	
\usepackage{amssymb}	
\usepackage{multicol}        
\usepackage{bm}		
\usepackage{pdflscape}	

\usepackage[T1]{fontenc}
\usepackage{ae,aecompl}

\usepackage{txfonts}

\title[Scaling relations of dwarf galaxies]{The scaling relations of early-type dwarf galaxies across a range of environments}

\author[S. J. Penny]{Samantha J. Penny$^{1,2, 3}$\thanks{Contact e-mail: \href{mailto:samantha.penny@port.ac.uk}{samantha.penny@port.ac.uk}}, Joachim Janz$^{4}$, Duncan A. Forbes$^{4}$, Andrew J. Benson$^{5}$ \newauthor and Jeremy Mould$^{4}$\\
$^{1}$ Institute of Cosmology and Gravitation, University of Portsmouth, Dennis Sciama Building, Burnaby Road, Portsmouth, PO1 3FX, UK\\
$^{2}$School of Physics, Monash University, Clayton, Victoria 3800, Australia\\
$^{3}$Monash Centre for Astrophysics, Monash University, Clayton, Victoria 3800, Australia\\
$^{4}$Centre for Astrophysics \& Supercomputing, Swinburne University, Hawthorn VIC 3122, Australia\\
$^{5}$Observatories of the Carnegie Institution for Science, 813 Santa Barbara Street, Pasadena, CA 91101, USA}

\date{Last updated 2015 May 22; in original form 2013 September 5}

\pubyear{2015}

\begin{document}
\label{firstpage}
\pagerange{\pageref{firstpage}--\pageref{lastpage}}
\maketitle

\begin{abstract}
We present the results of a Keck-\textsc{esi} study of dwarf galaxies across a range of environment: the Perseus Cluster, the Virgo Cluster, the NGC~1407 group, and the NGC~1023 group. 
Eighteen dEs are targeted for spectroscopy, three for the first time. 
We confirm cluster membership for one Virgo dE, and group membership for one dE in the NGC~1023 group, and one dE in the NGC~1407 group for the first time.
Regardless of environment, the dEs follow the same size-magnitude and $\sigma$-luminosity relation. 
Two of the Virgo dwarfs, VCC~1199 and VCC~1627, have among the highest central velocity dispersions ($\sigma_{0} = 58.4$~km~s$^{-1}$ and $49.2$~km~s$^{-1}$) measured for dwarfs of their luminosity ($M_{R}\approx -17$). 
Given their small sizes ($R_{e} < 300$~pc) and large central velocity dispersions, we classify these two dwarfs as compact ellipticals rather than dEs. 
Group dEs typically have higher mean dynamical-to-stellar mass ratios than the cluster dEs, with $M_{dyn}/M_{\star} = 5.1\pm0.6$ for the group dwarfs, vs. $M_{dyn}/M_{\star} = 2.2\pm0.5$ for the cluster sample, which includes two cEs.  
We also search for trends in $M_{dyn}/M_{\star}$ vs. distance from M87 for the Virgo Cluster population, and find no preference for dwarfs with high values of $M_{dyn}/M_{\star}$ to reside in the cluster outskirts vs. centre. 
\end{abstract}

\begin{keywords}
galaxies: dwarf -- galaxies:evolution
-- galaxies: kinematics and dynamics
\end{keywords}




\section{Introduction}

The total mass of a galaxy and its local environment are
fundamental parameters that determine its average properties and 
evolutionary history. 
For a dwarf galaxy, its total mass is generally dominated by 
its dark matter, with the mass in stars making a smaller overall contribution. 
In order to measure the total mass of a
dwarf galaxy, and probe its fraction of dark and stellar
matter,  a study of its dynamics is required. 
Outside of the Local Group where dwarfs can be resolved into their individual stars, such studies typically measure the motion of stars  within a projected radius
containing half of the total galaxy light (called the half-light
or effective radius R$_e$). 
Past studies of the dynamics of galaxies that are pressure-supported by random internal 
stellar motions (which are also typically old and cold gas-free systems) have ranged from 
the most massive elliptical galaxies to the lowest-luminosity dwarfs \citep[e.g.][]{2008MNRAS.386..864D,2008MNRAS.389.1924F,2010MNRAS.406.1220W,2011ApJ...726..108T,2011MNRAS.413.2665F,2011MNRAS.414.3699M,2014MNRAS.444.2993F}.

When examining galaxy kinematics, the stellar mass regime of $10^{7} - 10^{9}$~M$_{\odot}$ is particularly interesting as it represents a transition from 
the most massive dwarf spheroidal (dSph) to the lowest-mass dwarf elliptical (dE) galaxies. Spanning this entire mass range are the less extended compact ellipticals (cEs) and ultra-compact dwarfs (UCDs). 
For a given stellar mass, dEs, cEs and UCDs overlap in terms of their dynamical mass, though cEs and UCDs are more compact and have larger velocity dispersions than dEs/dSphs, so they appear as a distinct population in size-luminosity and $\sigma$-luminosity diagrams. 
Indeed, the compact cEs/UCDs are also referred to as ``true'' dwarf ellipticals in \citet{2012ApJS..198....2K}, given they extend the scaling relations of elliptical galaxies to fainter magnitudes. 
Their location in these diagrams is consistent with the idea that UCDs are the remnant nuclei or cores of stripped dE galaxies, and cEs are the remnants of more massive stripped objects such as spirals and low mass  ellipticals \citep[e.g.][]{2014MNRAS.443.1151N}.

Velocity dispersions and dynamical masses have been measured for a number of 
relatively high-mass dEs \citep[e.g.][]{2002AJ....124.3073G,2003AJ....126.1794G, 2009MNRAS.394.1229C, 2012A&A...548A..78T, 2014arXiv1410.1552T}, but such measurements for
dEs with dynamical masses $<10^{9}$~M$_{\odot}$ are relatively rare outside of the Local Group \citep[e.g.][]{2011MNRAS.413.2665F}. 
Obtaining kinematics for such low-mass dE galaxies  becomes difficult due to the combination of their low surface brightness and small velocity dispersions ($<30$~km~s$^{-1}$), necessitating high-resolution spectroscopy acquired with a fast telescope and spectrograph.

The high-mass dEs studied to date tend to be located in the Virgo Cluster, given its proximity and large population of such galaxies. 
A few studies have targeted dwarfs in other clusters including Fornax \citep[e.g.][]{2003A&A...400..119D}, and recently Perseus \citep{2014MNRAS.443.3381P}. 
Little is known however about the kinematics of dEs in group environments (with the exception of 3 dEs in the Local Group, see \citealt{2012AJ....144....4M} for a compilation of Local Group galaxy kinematics).

Investigating the dynamics and scaling relations of dEs in various environments may help to shed light on the dominance of different processes that effect their evolution. 
In groups, the dominant process is likely to be tidal interactions between galaxies or with the group potential itself \citep{1996Natur.379..613M,2005MNRAS.364..607M,2009Natur.460..605D}.
In the simulations of \citet{2006MNRAS.369.1021M}, late-type galaxies  in a Milky Way-like halo underwent strong tidal stripping, interactions, and ram pressure stripping, resulting in early-type dwarfs that are dark matter dominated in their central regions. 
Studies of galaxy dynamics are essential to identify these objects and measure the ratio of dark to luminous matter in their centres.

Tidal interactions and the stripping of low-mass galaxies can also lead to the formation of compact elliptical (cE) galaxies and ultra-compact dwarfs (UCDs). 
Compact ellipticals with masses comparable to dEs ($<10^{9}$~M$_{\odot}$) overlap in luminosity and velocity dispersion with dEs, though their sizes are much smaller ($R_{e} < 400$~pc). 
Continuing this trend to lower masses and sizes are UCDs, which overlap in stellar mass and luminosity with dSphs. 
In the simulations of tidal stripping by \citet{2013MNRAS.433.1997P},  UCDs with stellar masses  $\sim$ 10$^6$ M$_{\odot}$ and effective radii $\sim$ 10-50~pc are expected to form from low-mass dE progenitors. 
Such objects have been identified to be relatively common around M87 \citep{2011AJ....142..199B}. 
Thus by examining the kinematics of galaxies in the overlap between cEs and dEs, we can better understand the role of tidal interactions on low-mass galaxy evolution. 

Here we investigate the internal kinematics for a sample of 18 dwarf galaxies: 15 dEs and 3 possible cEs. 
The galaxies we examine have relatively low stellar masses of $\sim$10$^{8}$ ~M$_{\odot}$ to 10$^9$~M$_{\odot}$ and probe three different environments:
the spiral dominated NGC~1023 group, the fossil group-like NGC~1407 group, and the Virgo Cluster.  
The targeted dEs span a range of size ($ 0.27~\mathrm{kpc} < R_{e} < 1.93~\mathrm{kpc}$), covering the complete size range examined in the literature.  
They are among the lowest stellar mass dEs examined outside of the Local Group to date, with the lowest-mass dEs in our sample having stellar mass $\sim$$10^{8}$~M$_{\odot}$, comparable to the stellar mass of NGC~205, the brightest dE in the Local Group.
As well as central velocity dispersions, we present their sizes and luminosities. 
From these we calculate dynamical and stellar masses, and thus explore where such dE galaxies lie on key scaling relations including the sigma-luminosity and size-magnitude relations.

This paper is organised as follows. We describe our sample selection in Section~\ref{sec:sample}. 
Our observations are described in Section~\ref{sec:obs}, with the reduction of the spectroscopy covered in Section~\ref{sec:reduction}. 
Velocity measurements are presented in Section~\ref{sec:velocity}, along with photometry in Section~\ref{sec:photometry}. 
Our calculations of stellar and dynamical mass are presented in Section~\ref{sec:stellar}.  
We discuss our results in Section~\ref{sec:discuss}, and conclude in Section~\ref{sec:conclude}.

\section{Sample Selection}
\label{sec:sample}

Galaxies in the NGC~1023 group were selected from the catalogue of \citet{2009MNRAS.398..722T}. 
The three target galaxies have dE or intermediate-type morphology, and are brighter than $R = 15$, corresponding to $M_{R} = -14$ at the distance of the NGC~1023 group ($D=11.1$~Mpc, \citealt{2011AJ....142..199B}).  
NGC~1023\_11 and NGC~1023\_14 had their group membership confirmed in \citet{2009MNRAS.398..722T}, and we targeted NGC~1023\_18 for spectroscopy for the first time. 

We selected dwarfs in the NGC~1407 group  ($D=26.8$~Mpc) from the catalogue of \citet{2006MNRAS.369.1375T}. 
All targeted dwarfs   are brighter than $R=16.67$ ($M_{R} = -15.5 $), have dE morphologies, and have not previously had their central velocity dispersions measured. 
Three of the NGC~1407 dwarfs (NGC~1407\_13, NGC~1407\_47, and NGC~1407\_48) have their group membership confirmed in \citet{2006MNRAS.369.1375T}. 
NGC~1407\_36 and NGC~1407\_37 had their group membership confirmed in \citet{2006MNRAS.372.1856F}. 
We targeted NGC~1407\_43 for spectroscopy for the first time. 

Dwarf elliptical targets in the Virgo Cluster were selected from the ACS Virgo Cluster Survey \citep{2004ApJS..153..223C}, and the wide field imaging survey of \citet{2008AJ....135..380L}.  
The dwarfs were selected to have a surface brightness high enough  to obtain a spectrum with a signal-to-noise ratio S/N~$>10$ in less than 2~hr total integration time with the Echelle Spectrograph and Imager spectrograph \citep[\textsc{esi,}][]{2002PASP..114..851S} on the Keck~\textsc{ii} telescope. 
The majority of Virgo Cluster dwarfs targeted here had their cluster membership previously confirmed by the Sloan Digital Sky Survey \citep{2000AJ....120.1579Y}, but they do not have measured velocity dispersions in the literature. VCC~1151 is targeted for spectroscopy for the first time. 
The targets have a range of (projected) cluster-centric distance, from 0.28~Mpc to 1.35~Mpc from M87, the central galaxy in the Virgo Cluster. 
Five of the Virgo targets  were classified as E0 or E2 in the Virgo Cluster Catalogue \citep{1985AJ.....90.1681B}. 
Despite these classifications, these galaxies are targeted due to their low stellar mass ($<2\times10^{9}$~M$_{\odot}$), placing them well within the dwarf galaxy regime. 
Their sizes are furthermore intermediate between those of cEs and dEs ($ 270~\rmn{pc} < R_{e} < 850$~pc), making them interesting targets for the study of kinematics. 

The basic properties of the observed dwarfs are given in Table~\ref{table:photometry}. 
For simplicity, we adopt the names of Trentham \& Tully (2009) and Trentham et al.~(2006) for galaxies in the NGC~1023 and NGC~1407 groups. 
We take the names of the Virgo Cluster targets from the Virgo Cluster catalogue \citep{1985AJ.....90.1681B}. 
Assuming the targeted dwarfs follow the $\sigma$-luminosity relation, they are expected to have central velocity dispersions comparable to, or exceeding, the \textsc{esi} instrumental resolution (see Section~\ref{sec:obs}).

These data are supplemented with three dwarfs in the Perseus Cluster: CGW38, CGW39, and SA0426-002, first presented in \citet{2014MNRAS.443.3381P}. 
Perseus is a richer cluster than Virgo, providing a more extreme environment in which to study galaxy evolution. 
These dwarfs were observed with an identical set-up to the Virgo targets in this work, using \textsc{esi}  and $0.5$~arcsec slit. 
 The reduction of these data are described in \citet{2014MNRAS.443.3381P}.

\begin{table*}
\caption{Properties of the dE galaxy sample. The galaxy types are from taken from the NASA Extragalactic Database (NED), with the exception NGC~1023\_18, which is classified in Trentham \& Tully (2009). The distances of the two galaxy groups are taken from \citet{2011AJ....142..199B}.
Where available, surface brightness fluctuation distances are provided for dEs in the Virgo Cluster, otherwise 
the surface brightness fluctuation distance to M87 from \citet{2009ApJ...694..556B}  is used. The $R$-band magnitudes are described in Section~\ref{sec:photometry}. The half-light radii for dwarfs in the NGC~1023 and NGC~1407 groups are measured in this work, and taken from Janz \& Lisker (2008) for the Virgo Cluster objects. 
The effective radius in kpc is calculated using the distance provided in the table. 
The last column lists the S\'ersic index $n$ of our surface brightness fits.}\label{table:size}
\begin{tabular}{lccccccc}
\hline
Galaxy & Other name & Type & Dist. & R &  R$_e$  & R$_e$ & $n$ \\
& & &(Mpc) & (mag)  & ($^{''}$) & (kpc) & \\  
\hline
NGC~1023\_11 & UGC~2165 &dE    & 11.1 & 13.70 &    17.6  	    &	    0.95 & 1.0 \\
NGC~1023\_14 &UGC~1807  &Im    & 11.1 & 14.38 &    22.8 	    & 	    1.23  & 1.4 \\
NGC~1023\_18 & \ldots &dE/I  & 11.1 & 14.92 &    19.4 	    &      1.04 & 1.6 \\
\hline					 					   
NGC~1407\_13 & ESO~548-G79 &dE    & 26.8 & 13.03 &    14.9	    & 	    1.93 & 1.9 \\
NGC~1407\_36 &LEDA074838  &dE    & 26.8 & 15.33 &    8.5	    &	    1.11 & 1.4  \\
NGC~1407\_37 &LSBG~F548-012 &dE    & 26.8 & 15.52 &    11.6 	    &	    1.50 & 1.1 \\
NGC~1407\_43 & LSBG~F549-023 &dE    & 26.8 & 16.40 &    8.4   	    &	    1.10 & 0.9 \\
NGC~1407\_47 & LSBG~F549-038 &dE    & 26.8 & 16.70 &    6.3 	    &  	    0.83 & 1.3\\
NGC~1407\_48 & LSBG~F548-026 &dE    & 26.8 & 16.67 &    6.7	    & 	    0.87  & 0.9\\
\hline
VCC~50  &  \ldots & dE2  & 16.7 & 14.89  & 12.5    &	0.99  & 0.9  \\    
VCC~158  & UGC~07269 & dE3  & 16.7 & 14.61    & 20.9    &	1.66  & 1.2  \\    
VCC~538  & NGC~4309A & E0    &  $23.0\pm0.9$ & 14.90 & 4.8     &	0.54  & 2.3  \\    
VCC~1151 & NGC~4472 DW02 & dE0  & 16.7 & 15.29& 17.5    &	1.39  & 1.2  \\    
VCC~1199 &  \ldots  & E2    & $16.3\pm1.5$ & 14.91 & 3.4   &	0.27  & 2.9   \\    
VCC~1440 & IC~0798 & E0   & $16.1\pm0.6$ & 13.55  & 6.9      &	0.54  & 4.8   \\    
VCC~1627 &  \ldots  & E0   & $15.6\pm1.6$ & 13.92 & 3.7	     &	0.28  & 2.0   \\    
VCC~1896 &  \ldots  & dSB0 & 16.7 & 13.82  & 14.6 &	1.16  & 1.2  \\    
VCC~1993 &  \ldots  & E0   & $16.6\pm0.5$ & 14.28 & 10.7   &	0.86   & 4.6  \\    
\hline
\end{tabular}
\\
\label{table:photometry}
\end{table*}

\section{Data}
\label{sec:data}

\subsection{Observations}
\label{sec:obs}

The dwarf elliptical galaxies were observed using the Echelle
Spectrograph and Imager  on the Keck \textsc{ii} 10~m telescope on the
nights of 2013 September 25, 26, 2013 March 3, 4, 
and 2012 November 5, 6, 7. 
Each galaxy was observed in
high resolution
echelle mode giving a useful 
wavelength range of
$\sim$4000 to 10,000 \AA. The pixel scale varies from 0.12~arcsec~pixel$^{-1}$
in the blue to 0.17~arcsec~pixel$^{-1}$ in the red 
across the 10 echelle orders. The slit width was either 
0.5~arcsec or 0.75~arcsec (see Table 2), with a slit length of 
20~arcsec. This provided a resolving power of $R=8000$ or $R=5400$, and instrumental resolutions of 15.8~km~s$^{-1}$ and 23.7~km~s$^{-1}$, for the $0.5$~arcsec and $0.75$~arcsec slits respectively. 
Generally the slit was aligned to parallactic angle, otherwise for dwarfs that deviated strongly from circular isophotes, we aligned the slit with the dwarf's major axis. 

Three or more single exposures were taken of each galaxy 
with the  total exposure times  listed in Table 2. 
The individual exposures were  combined 
with an average sigma clipping algorithm in IRAF.
We obtained spectra for several standard stars on each observing run
using the same instrument settings as for the science data. These spectra were 
used as templates to determine velocity information for the dwarf galaxies 
(see below for details).

\begin{table}
\caption{Observing parameters for the dEs targeted in this work with Keck \textsc{esi}.}
\begin{tabular}{lccc}
\hline
Galaxy  & Seeing & Exp. time & Slit width\\
       &  (arcsec) &  (min) & (arcsec)\\
\hline
NGC~1023\_11  & 0.9 & 30 & 0.75\\
NGC~1023\_14  & 0.7 & 120 & 0.75\\
NGC~1023\_18  & 0.7 & 80 & 0.75\\
\hline
NGC~1407\_13  & 0.7 & 15 & 0.75\\
NGC~1407\_36  & 0.7 & 60 & 0.75\\
NGC~1407\_37  & 0.7 & 80 & 0.75\\
NGC~1407\_43 & 0.5 & 100 & 0.75\\  
NGC~1407\_47 & 0.5 & 100 & 0.75\\
NGC~1407\_48 & 0.7 & 80 & 0.75\\
\hline
VCC~50  & 0.95 & 120 & 0.50\\
VCC~158 & 0.95 & 100 & 0.50\\
VCC~538  & 1.2 & 100 & 0.50\\
VCC~1151 & 0.95 & 80 & 0.50\\
VCC~1199  & 1.2 & 30 & 0.50\\
VCC~1440 & 1.2 & 80 & 0.50\\
VCC~1627  & 1.2 & 100 & 0.50\\
VCC~1896 & 0.95 & 30 & 0.50\\
VCC~1993 & 1.2 & 80 & 0.50\\
\hline
\end{tabular}
\end{table}

\subsection{Spectroscopic Data Reduction}
\label{sec:reduction}

\begin{figure*}
\begin{center}
\includegraphics[width=0.97\textwidth]{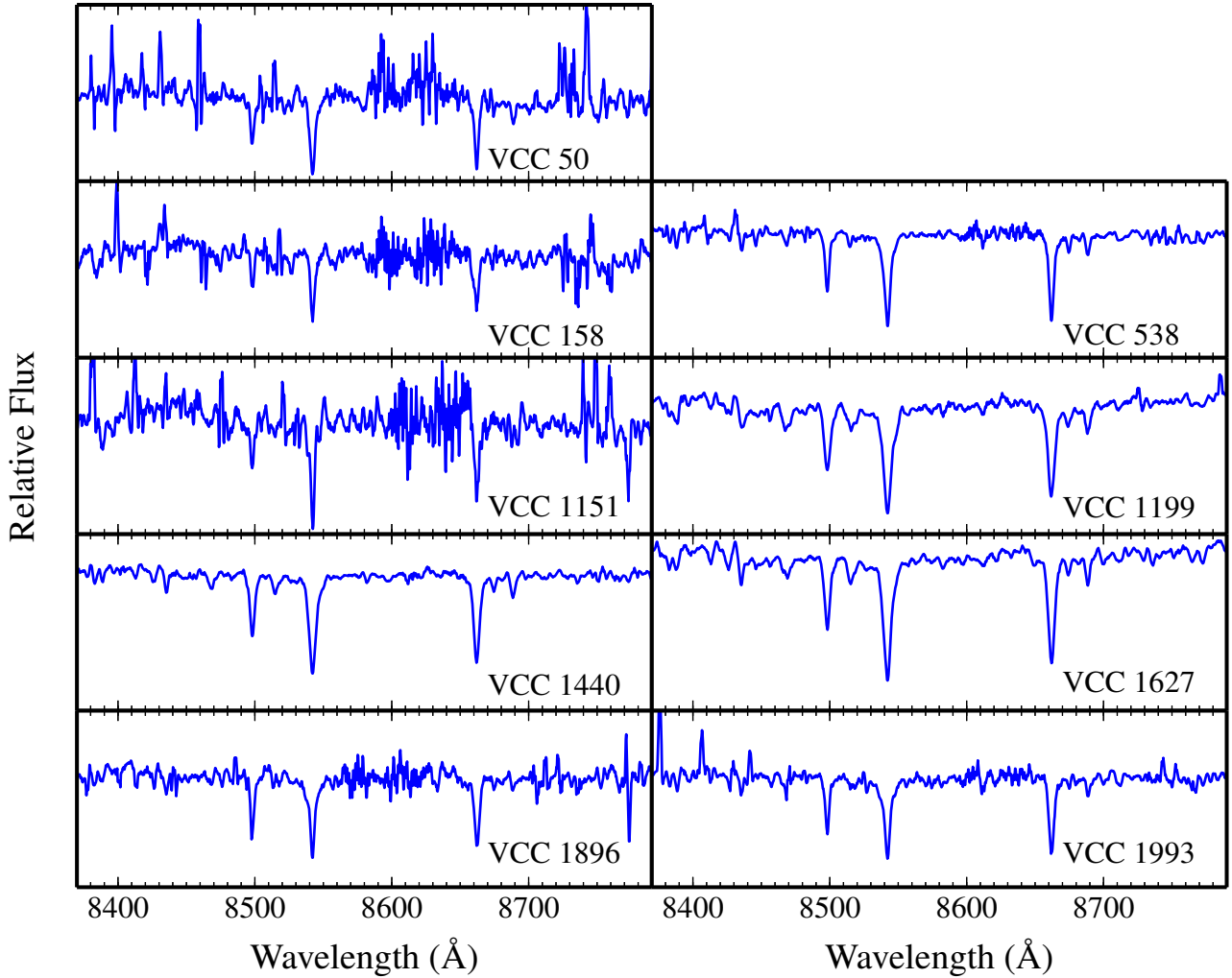}
\caption{\textsc{esi} spectra for the Virgo dwarfs presented in this study, showing the CaT feature. The spectra have been shifted to rest-frame wavelengths, and have been smoothed. Due the presence of skylines, not all CaT lines could be used in all kinematic fits. We also use the not shown H$\alpha$, Fe and Mg lines when available.}\label{virg_specs}
\end{center}
\end{figure*}

\begin{figure*}
\begin{center}
\includegraphics[width=0.97\textwidth]{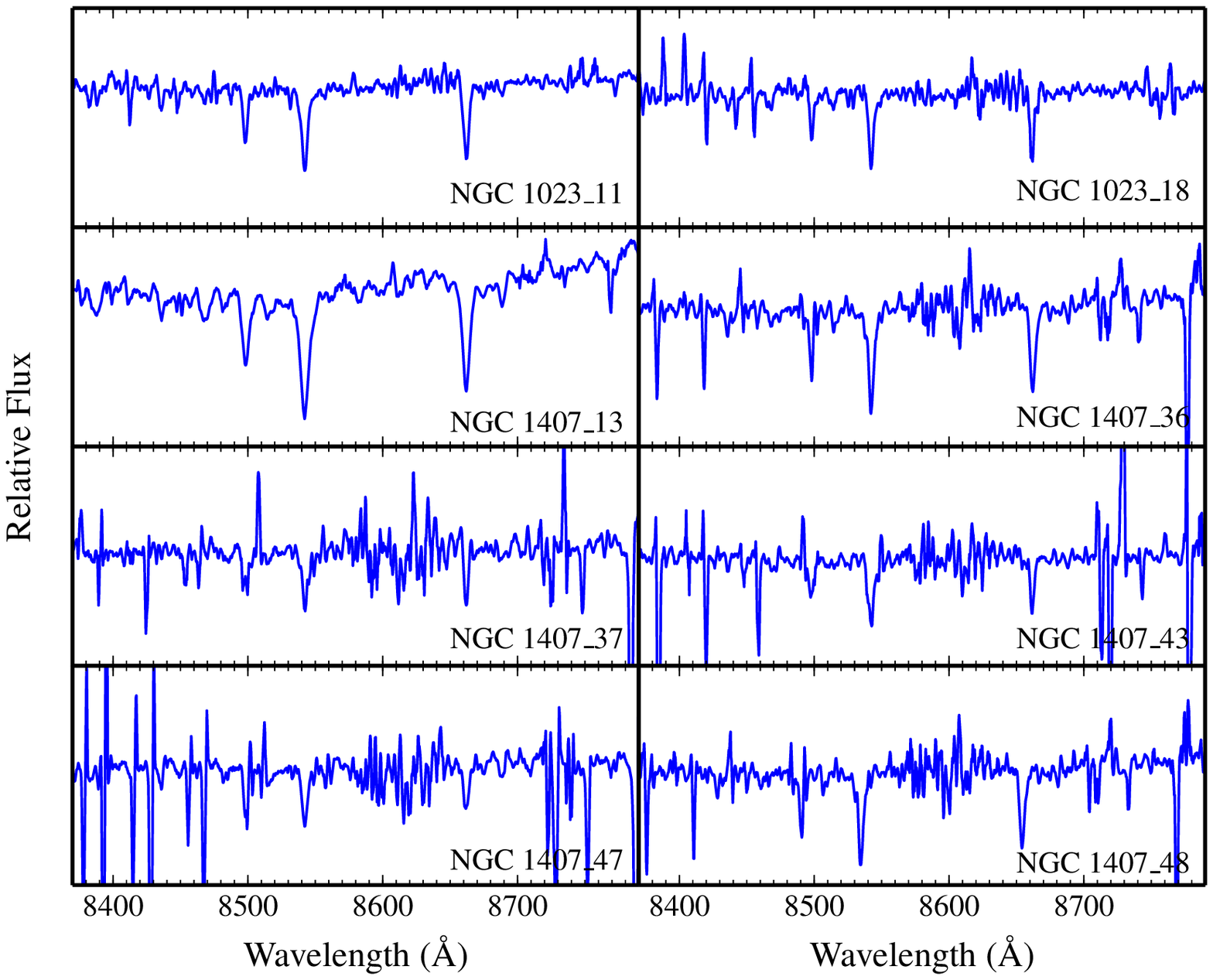}
\caption{Same as Fig.~\ref{virg_specs}, but for dwarfs in the NGC~1023 and NGC~1407 groups. NGC~1023\_14 is not included due to poor S/N in the CaT region. }\label{group_specs}
\end{center}
\end{figure*}

Basic data reduction for the \textsc{esi} spectroscopy was carried out using tasks within IRAF.  
Individual calibration frames such as bias, arcs and 
internal flat fields were combined to
create master frames. 
The science frames, each of the
same exposure time, were average combined in 2D.
The individual science frames did not
require shifting as the spatial alignment of the spectra was
within 1 pixel from frame to frame. 
An average sigma clipping was used to reject cosmic rays. 

Tracing and rectifying the spectra, wavelength calibration,
extraction of 1D spectra in various apertures and  sky
subtraction were all performed using the \textsc{MAKEE} program written by
T. Barlow. 
The trace was carried out using a bright standard
star, which gave residuals of $\le 0.5$~pixel for the orders of interest.
The sky was measured from the edges of the
slit, furthest from the galaxy centre. 
Although some faint 
galaxy light 
may be contained in the sky apertures, there was no indication of the CaT
absorption lines in the sky spectra, so that this appears to be a very
small effect.

For five Virgo galaxies, VCC~538, VCC~1199, VCC~1440, VCC~1627 and VCC~1993, the spectra extend further than  the 
seeing profile. 
For these we extracted additional independent apertures either side of the galaxy centre, in addition to a central aperture of $\pm$4 pixels ($\sim1.2$~arcsec width).
The midpoint of the apertures correspond to 1.4, 2.6, and 4.4~arcsec from the galaxy centre. 
The size of the off-centre extraction apertures was designed to achieve a similar S/N in each extraction independent of radius. 

\begin{figure}
\begin{center}
\includegraphics[width=0.47\textwidth]{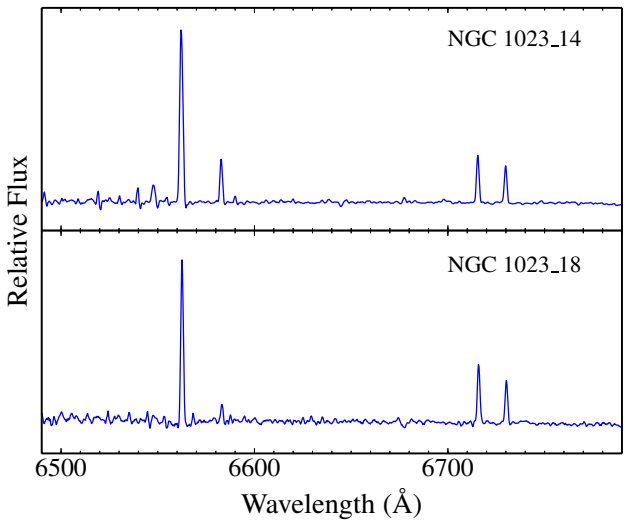}
\caption{\textsc{esi} spectra for two dwarfs in the NGC~1023 group with ongoing star formation, showing the emission lines in the vicinity of H$\alpha$. }\label{haspecs}
\end{center}
\end{figure}

We plot  central spectra for all dwarfs examined in this study to highlight the quality of our data. 
Spectra of the Virgo dwarfs are shown in Figure~\ref{virg_specs}, and spectra for the NGC~1407 and NGC~1023 dwarfs  in Figure~\ref{group_specs}. 
The CaT feature is plotted for all objects, excluding NGC~1023\_18, where the S/N was insufficient for the CaT feature to be discernible against the sky lines.  
Due to the radial velocities of the dwarfs examined here, individual CaT lines are frequently shifted into regions of the spectra strongly affected by sky lines. 
We therefore smoothed the spectra for plotting purposes. 
For the science analysis, all spectra were unsmoothed/unbinned. 
To highlight that two of the NGC~1023 dwarfs are star forming, we display their spectra in the  H$\alpha$ region in Figure~\ref{haspecs}. For both NGC~1023\_14 and NGC~1023\_18, the H$\alpha$, [N\textsc{ii}] and [S\textsc{ii}] lines are clearly seen in emission, indicating ongoing star formation. 

\subsection{Velocity Measurements}
\label{sec:velocity}

Kinematics were measured for each extracted aperture using the Penalized Pixel-Fitting method (pPXF, Cappellari \& Emsellem, 2004).  
The code broadens the spectra of the standard stars until they match the observed galaxy. We fitted the kinematics for several regions of each galaxy spectrum: the calcium triplet  feature (CaT; 8498, 8542, 8662~\AA), the H$\alpha$ (6563~\AA) line, the H$\beta$ line (4861~\AA), the Na line (5893~\AA), and the Fe and Mg features (5100-5400 \AA).  
Where necessary, individual CaT lines that are blended with sky lines were excluded from the pPXF fit.  The resulting heliocentric velocities ($cz$) and velocity dispersions ($\sigma$) are listed in Table~\ref{table:velocities}.

NGC~1023\_14 and NGC~1023\_18 have H$\alpha$ in emission, and we were unable to use this line for the determination of the  velocity dispersions (our template stars only have absorption features). We furthermore excluded the H$\beta$ line for these objects, as this feature is in emission for NGC~1023\_18, and partially infilled for NGC~1023\_14.  Instead, the dwarfs' central velocity dispersions are calculated using the CaT, Na, and Fe/Mg features only. For NGC~1023\_14, the CaT and Mg/Fe features do not have high enough S/N for the measurement of its velocity dispersion, so we are only able to provide a redshift for this object.

\begin{figure}
\includegraphics[width=0.48\textwidth]{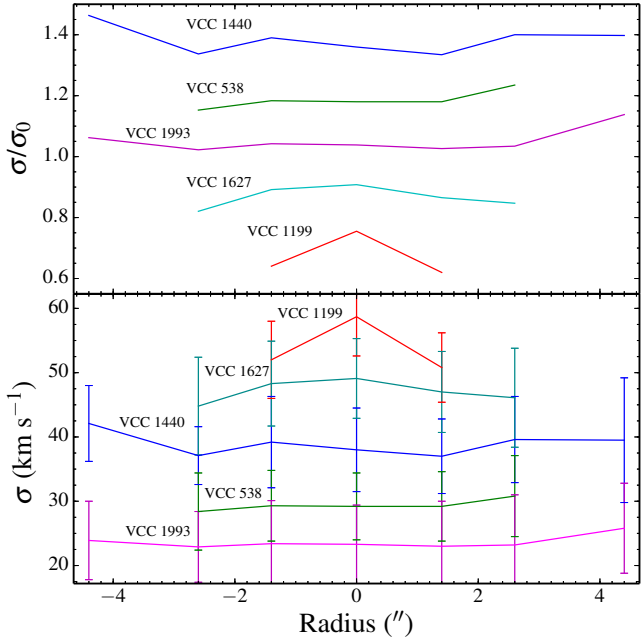}
\caption{Velocity dispersion profiles.  {\it Top panel:} velocity dispersions for the 5 Virgo dEs with spatially resolved spectroscopy, out to $4\farcs4$ from their centres, normalised  to their central velocity dispersion $\sigma_{0}$ at $R_{e}/8$.  An arbitrary offset has been applied to the profiles to separate them. 
Other than VCC~1199, there is no clear evidence for central peaks in the velocity dispersion profiles. {\it Bottom panel:} unnormalised velocity dispersion profiles. Error bars are also plotted. The central values are the 1.2~arcsec aperture values, and have not been corrected to $R_{e}/8$. }\label{dispprofs}
\end{figure}

For consistency with the literature, we corrected our aperture values to central values within $R_{e}/8$ \citep[e.g.][]{1995MNRAS.276.1341J}, following \citet{2006MNRAS.366.1126C}: 

\begin{equation}
\sigma_{0} = \sigma_{ap}\left(\frac{8 R_{ap}}{R_{e}}\right)^{-0.066}\label{eqn:mass}
\end{equation}

\noindent where $\sigma_{0}$ is the central velocity dispersion, $\sigma_{ap}$ is the velocity dispersion in the central $1\farcs2$ aperture, $R_{ap}$ is the radius of the extraction aperture, and $R_{e}$ is the effective radius of the dwarf. 
The extraction apertures have radii  0.45~arcsec and 0.55~arcsec for the 0.5 and 0.75~arcsec slits respectively, calculated using the formula $2~R_{ap} \approx 1.025 \times 2(xy/\pi)^{1/2}$ from \citet{1995MNRAS.276.1341J}.
These central values $\sigma_{0}$ are provided in Table~\ref{table:masses}.

The majority of the comparison samples we utilise in this work correct their values of $\sigma$ to $R_{e}/8$, though \citet{2014ApJS..215...17T} and \citet{2014MNRAS.439..284R} correct their values to 1~$R_{e}$. 
These values of $\sigma_{e}$ will differ from $\sigma_{0}$ by a few km~s$^{-1}$, within typical error bars, so we do not correct these comparison samples to $R_{e}/8$.

\begin{table}
\caption{Velocity measurements for the dEs targeted in this work. $\sigma_{ap}$ is the velocity dispersion measured from the full slit aperture.
The values of $\sigma_{ap}$ are calculated from 3 regions: CaT, H$\alpha$, and the Mg, Fe lines. 
Heliocentric corrections have been applied to all recessional velocities.
The velocity dispersions for VCC~50, VCC~158, and NGC~1407\_47 are flagged as upper limits due to their spectra having low S/N $<10$ and measured velocity dispersions comparable to the instrumental resolution for their observational setup.}
\begin{tabular}{lcc}
\hline
Galaxy Name & Galaxy $\sigma_{ap}$ & Recession Velocity \\
 & (km s$^{-1}$) & (km s$^{-1}$) \\
\hline
NGC~1023\_11 & $31.6\pm5.4$ & $ 699\pm8$  \\
NGC~1023\_14 &  \ldots & $594 \pm 24$ \\
NGC~1023\_18 & $22.5 \pm 7.9$ & $ 438 \pm 8$ \\
\hline
NGC~1407\_13 & $60.8\pm5.9$ & $2002\pm7$ \\
NGC~1407\_36 & $25.0\pm5.5$ &  $1726\pm9$  \\
NGC~1407\_37 &$ 28.1\pm8.0$ & $1476\pm14$  \\
NGC~1407\_43 & \ldots & $1684\pm7$   \\
NGC~1407\_47 & $<24.3\pm5.2$ & $1349\pm12$ \\
NGC~1407\_48 & $26.5 \pm 5.1$ &  $1372\pm8$ \\
\hline
VCC~50 & $<13.5 \pm 5.9$ & $1220 \pm 11$ \\
VCC~158 & $<16.1 \pm 5.4$  & $1083\pm6$  \\
VCC~538 & $ 29.2  \pm5.2$ & $743\pm5$ \\
VCC~1151 & \ldots & $630\pm8$ \\
VCC~1199 & $58.7  \pm6.1$ & $1395\pm8$  \\
VCC~1440 & $38.0 \pm6.5$  & $431\pm5$  \\
VCC~1627 & $49.1 \pm6.2$ & $294\pm5$ \\
VCC~1896 & $22.4 \pm 6.5$ & $1867\pm8$  \\
VCC~1993 & $23.3 \pm6.1$  & $875\pm6$ \\
\hline
\end{tabular}
\label{table:velocities}
\end{table}

For the five Virgo dwarfs resolved beyond the seeing profile, we derive the velocity dispersion $\sigma$ in several radial bins out to radii of $4.4$~arcsec, $\sim$300~pc at the distance of the Virgo Cluster. 
The velocity dispersions in each of the radial bins are provided in Table \ref{table:sigma}.
We plot the velocity dispersion profiles for these five dEs in Figure~\ref{dispprofs}.
Within a radius of  $300$~pc, the velocity dispersions profiles of these objects are nearly flat, with no evidence for central peaks or dips, similar to the results of \citet{2003AJ....126.1794G,2011MNRAS.413.2665F,2011A&A...526A.114T} for dEs, and those for the Local Group dSphs.
To the same radius of $4.4$~arcsec, we do not find any rotation to within $\pm$10 km~s$^{-1}$ of the central value. 
Therefore, we classify these five dwarfs as slow rotators. 
More extended spectroscopy beyond their effective radii may uncover rotation in the outer regions of these objects.
Nevertheless, the kinematics of these five dwarfs are dominated by random motions rather than rotation.

\begin{table*}
\caption{Spatially resolved velocity dispersion measurements for the five dwarfs with spectra extending beyond the seeing profile. For each galaxy we list the velocity dispersion  in radial bins about the central 
aperture. The central aperture is 1.2~arcsec wide. The $\sigma$ values are aperture values, i.e. they have not been corrected to $R_{e}/8$. }
\begin{tabular}{lccccccc}
\hline
  \multicolumn{1}{c|}{Galaxy}  &   \multicolumn{7}{c|}{Velocity dispersion in radial bin }    \\
 & $-4\farcs4$ & $-2\farcs6$ & $-1\farcs4$ & $0\arcsec$ & $1\farcs4$ & $2\farcs6$ & $4\farcs4$ \\
 & (km s$^{-1}$) & (km s$^{-1}$) & (km s$^{-1}$)&  (km s$^{-1}$) & (km s$^{-1}$) & (km s$^{-1}$) & (km s$^{-1}$)\\
 \hline
VCC~538 & $\ldots$ & $28.4\pm6.0$ & $29.3\pm5.5$ & $29.2\pm5.2$ & $29.2\pm5.4$ & $30.8\pm6.3$ & $\ldots$ \\
VCC~1199 & $\ldots$&  $\ldots$& $52.0\pm 6.0$ & $58.7\pm 6.1$ & $50.8\pm 5.4$ & $\ldots$& $\ldots$\\
VCC~1440 & $42.1\pm5.9$ &$37.1\pm4.5$ & $39.2\pm7.1$ & $38.0\pm6.5$ &$37.0\pm5.8$ & $39.6\pm6.7$& $39.5\pm9.7$ \\
VCC~1627 & $\ldots$& $44.8\pm7.6$ & $48.3\pm6.6$ & $49.1\pm6.2$ & $47.0\pm6.3$ & $46.1\pm7.7$ & $\ldots$\\
VCC~1993 & $23.9  \pm 6.1$ &  $22.9\pm0.5 $&  $23.4\pm6.7$ & $23.3\pm6.1$ & $23.0\pm7.0$ & $23.2 \pm 7.8$ & $25.8\pm7.0$  \\
\hline
\end{tabular}
\\
\label{table:sigma}
\end{table*}

\subsubsection{Signal-to-noise effects}
\label{sec:sn}

Several of our spectra have velocity dispersions approaching the instrumental resolution of Keck-\textsc{esi}.
 Velocity dispersion measurements are less accurate for low S/N spectra than for those with high S/N, such that for noisy galaxy spectra, the measured velocity dispersion typically approaches the instrumental resolution when the true $\sigma$ is low.  
 Furthermore, the reliability of the measured $\sigma$ decreases.
 For example,  \cite{2014ApJS..215...17T} measure velocity dispersion differences $>10$~km~s$^{-1}$ between their spectra with minimum S/N of 10, and those with previous measurements in the literature.

Six of our galaxies have velocity dispersions approaching the instrumental resolution. VCC~50 and VCC~158 have $\sigma \approx 15$~km~s$^{-1}$, the instrumental resolution of \textsc{esi} using the $0.5''$ slit. For observations taken with the $0.75''$ slit, NGC~1023\_18, NGC~1407\_36, NGC~1407\_47, and NGC~1407\_48 have $\sigma \approx 24$~km~s$^{-1}$, again comparable with the instrumental resolution. NGC~1407\_36 has S/N~$=15$ in the CaT region, sufficiently high that its velocity dispersion can be reliably measured. For NGC~1407\_48, while the CaT region is contaminated by sky lines, the S/N in the region of the H$\alpha$ line \textsc{esi} spectra is high (S/N$\sim$11). Thus we expect the value of $\sigma_{ap} = 26.5$~km~s$^{-1}$ we determine for this galaxy to be robust. 

The remaining galaxies with velocity dispersions comparable to the instrumental resolution have $S/N < 10$, and our $\sigma$ values should therefore be considered as upper limits on the true velocity dispersions of these objects. 
NGC~1407\_47 has S/N$=8$ in the CaT region, and $\sigma_{ap}=24.3$~km~s$^{-1}$, comparable to the instrumental resolution using the $0.75''$ slit.
VCC~50, and VCC~158 having S/N~$7$ and 8 respectively, and $\sigma$ values similar to the \textsc{esi} resolution of  15.8~km~s$^{-1}$ using the $0.5''$ slit. 
Given that the true $\sigma$ values for VCC~50, VCC~158 and NGC~1407\_47 are comparable to the instrumental resolution, they are likely over-estimated.  We therefore consider their dynamical masses to be slightly over-estimated.

\subsection{Size measurements and photometry}
\label{sec:photometry}

We measured the sizes of the dwarf galaxies in the NGC~1023 and NGC~1407 groups  on archival CFHT images \citep{2006MNRAS.369.1375T,2009MNRAS.398..722T}. 
The images were first sky subtracted, and masks for fore- and background objects created.
Subsequently, we extracted the light profiles using the \textsc{ellipse} task in \textsc{iraf} \citep{1987MNRAS.226..747J}, with the ellipticity and position angle  as free fitting parameters and logarithmic steps in semi-major axis.
S\'ersic functions were fitted to the light profiles using a Levenberg-Marquardt algorithm, taking into account the errors given by the ellipse task, and excluding the inner 2.5~arcsec from the fit in order to avoid being affected by the seeing and possible nuclei. 
The half-light semi-major axes ($a_{e}$) were obtained as one of the parameters of the fitting function.
 Finally, the effective radii were determined as the geometric mean of the semi-major and minor axes, i.e.\ $R_e=a_{e} \sqrt{b/a}$, with the axis ratio $b/a$ averaged around the half-light semi-major axis.

For the Virgo galaxies we use the $R_e$ measurements of \citet{2008ApJ...689L..25J}. These were obtained in a non-parametric, but homogeneous manner. 
 VCC~1199 is not included in the analysis of  \citet{2008ApJ...689L..25J}, and we therefore adopt its size from \citet{2006ApJS..164..334F}. 
 The S\'ersic indices presented in Table~\ref{table:photometry} were also obtained from these two works. 
 
The $R$-band magnitudes for the galaxies in the  NGC~1023 and NGC~1407 groups were taken from \citet{2009MNRAS.398..722T} and \citet{2006MNRAS.369.1375T}, respectively. 
No errors were provided for the photometry of galaxies in both groups in the literature, though uncertainties in their distances will dominate the error budgets of their absolute magnitudes, which are used in future calculations of galaxy luminosity in Section~\ref{sec:stellar}.
We assume errors of 5~per~cent on their photometry in later calculations of stellar mass. 
For the Virgo galaxies we converted the $r$-band values of \citet{2008ApJ...689L..25J} to the $R$-band using the conversion of \citet{2002AJ....123.2121S}, and the  $(g-r)$ colours of the galaxies \citep{2009ApJ...696L.102J}. The $(g-r)$ colour was selected as it uses the highest S/N bands of SDSS imaging, and have typical errors $<0.03$~mag. 
For VCC~1199, we convert the $g$-band value of \citet{2010ApJS..191....1C} to $R$ using its $(g-r)$ colour.

We investigate the robustness of our calculated $R$-band magnitudes by comparing them to the literature.  
The adopted $(g-r)$ colours of our dwarfs agree very well with those of \citet{2010ApJS..191....1C}, who compared their $g$-band photometry to that of \citet{2008ApJ...689L..25J}. They found a scatter $\sigma=0.09$~mag around a 1:1 relation between their measurements and those of   \citet{2008ApJ...689L..25J}. 
We confirm this agreement by calculating $R$ band magnitudes for our dwarfs from the $g$-band photometry and $(g-r)$ colours of  \citet{2010ApJS..191....1C} for the 4 Virgo dwarfs present in their photometry catalogue. The recalculated $R$-band magnitudes agree to within 0.1~mag of our values for all 4 galaxies.

We also compare our calculated values of $R$ to \citet{2014ApJS..215...22K}, converting their $r$-band values to $R$. Eight of the Virgo dEs in our sample are also found in their catalogue. The $R$-band values agree to within 0.3~mag for VCC~50, VCC~538, VCC1199, VCC~1440, VCC~1627, and VCC~1896, though are typically fainter than our calculated values. For VCC~158 and VCC~1993, the values we calculate from the photometry of \citet{2014ApJS..215...22K} are all fainter by $>0.3$~mag than the values we present from the photometry of \citet{2008ApJ...689L..25J} and \citet{2009ApJ...696L.102J}. Our photometry for these two galaxies agrees more closely with that presented by \citet{2010ApJS..191....1C}. We find for VCC~158 and VCC~1993, the effective radii presented in \citet{2014ApJS..215...22K} are smaller than those of \citet{2008ApJ...689L..25J} by $\sim20$~per~cent, resulting in the offset to fainter magnitudes when compared to our values of $R$. 
The combined effect of using these fainter magnitudes and smaller sizes could reduce their dynamical-to-stellar mass ratios by $\sim20$~per~cent.

Absolute magnitudes and physical sizes for the dwarfs in the NGC~1023 and NGC~1407 groups are calculated using the distances to their central galaxies. 
Where available, distances from surface brightness fluctuations from \citet{2009ApJ...694..556B} and \citet{2010ApJ...724..657B} are used in the calculation of the absolute magnitudes and physical sizes of the Virgo dEs, otherwise the distance to M87 (16.7~Mpc) is used. 
The adopted \textit{R}-band photometry for all galaxies is given in Table~\ref{table:photometry}.

\section{Mass estimates}
\label{sec:stellar}

\begin{table*}
\caption{Stellar and dynamical masses for dEs in four environments: the NGC~1023 group, the NGC~1407 group, the Virgo Cluster, and the Perseus Cluster. Stellar masses were calculated using the \citet{2005MNRAS.362..799M} models assuming a Kroupa IMF, a fixed metallicity [Z/H]$= -0.33$, and three mean stellar ages: 10~Gyr, 5~Gyr, and 3~Gyr. 
All values are determined out to 1~$R_{e}$, and the stellar and dynamical masses are therefore half-light masses. 
The values of $M_{R}$, $M_{\star}$ and $M_{dyn}$ are calculated using the distances listed in Table~\ref{table:size}.
 Also included are three dEs in the Perseus Cluster with \textsc{esi} spectroscopy first presented in \citet{2014MNRAS.443.3381P}, at an assumed distance of 70~Mpc. }\label{table:masses}
\begin{tabular}{lccccccccc}
\hline
  \multicolumn{1}{c|}{Galaxy} &
  \multicolumn{1}{c|}{$M_{R}$} &
  \multicolumn{1}{c|}{$\sigma_{0}$} &
  \multicolumn{1}{c|}{$M_{dyn}$} &
  \multicolumn{1}{c|}{$M_{\star}$ (10~Gyr)}&
  \multicolumn{1}{c|}{$M_{\star}$ (5~Gyr)} &
  \multicolumn{1}{c|}{$M_{\star}$ (3~Gyr)} &
  \multicolumn{1}{c|}{$M_{dyn}/M_{\star}$} &
  \multicolumn{1}{c|}{$M_{dyn}/M_{\star}$} &
  \multicolumn{1}{c|}{$M_{dyn}/M_{\star}$} \\
     &  (mag) & (km~s$^{-1}$ )&  ($\times10^{8}$~M$_{\odot}$) & ($\times10^{8}$~M$_{\odot}$) & ($\times10^{8}$~M$_{\odot}$)&  ($\times10^{8}$~M$_{\odot}$)& (10~Gyr) & (5~Gyr) & (3~Gyr)\\
\hline
  NGC~1023\_11 & $-16.53 \pm 0.21$ & 34.6 & $17.2   \pm 4.3$   & $3.6   \pm 0.6$ & $2.1 \pm 0.4$ & $1.3 \pm 0.4$ & $4.8 \pm 1.5$ & $8.3 \pm 3.3$ & $12.9 \pm 4.8$\\
  NGC~1023\_14 & $-15.85\pm 0.21$ & \ldots &  \ldots & $1.9\pm0.4$ & $1.1\pm0.2$ & $0.7\pm0.2$ &  \ldots &  \ldots &  \ldots\\  
NGC~1023\_18 & $-15.31\pm 0.21$  & 23.8 & $8.9     \pm 4.3$   & $1.2   \pm 0.3$ & $0.7  \pm 0.2$ & $0.4 \pm 0.2$ & $7.7 \pm 4.3$ & $13.2 \pm 9.1$ & $20.5 \pm 10.4$\\
\hline
  NGC~1407\_13 & $-19.11 \pm 0.26$ & 65.9 & $126.7 \pm 22.3$ & $38.4 \pm 7.9$ & $22.4 \pm 5.3$ & $14.4 \pm 4.7$ & $3.3 \pm 0.9$ & $5.7 \pm 2.2$ & $8.8 \pm 2.5$\\
  NGC~1407\_36 & $-16.81 \pm 0.26$ & 26.1 & $11.4   \pm 3.7$   & $4.6  \pm 1.0$ & $2.7 \pm 0.6$ & $1.7\pm 0.6$ & $2.5 \pm 0.9$ & $4.2 \pm 2.0$ & $6.6 \pm 2.6$\\
  NGC~1407\_37 & $-16.62 \pm 0.26$ & 29.9 & $20.3   \pm 8.1$   & $3.9   \pm 0.8$ & $2.3 \pm 0.5$ & $1.5 \pm 0.5$ & $5.2 \pm 2.3$ & $9.0 \pm 4.8$ & $13.9 \pm 4.3$\\
  NGC~1407\_43 & $-15.74 \pm 0.26$ &  \ldots &  \ldots & $1.7\pm0.4$ & $1.0\pm0.2$ & $0.7\pm0.1$ &  \ldots &  \ldots &  \ldots\\  
NGC~1407\_47 & $-15.44 \pm 0.26$ & $<24.9$ & $<7.8     \pm 2.5$   & $1.3   \pm 0.3$ & $0.8 \pm 0.2$ & $0.5 \pm 0.2$ & $<5.9 \pm 2.3$ & $<10.2 \pm 4.9$ & $<15.9 \pm 10.1$\\
  NGC~1407\_48 & $-15.47 \pm 0.26$ & 25.2 & $8.4     \pm 2.6$   & $1.3   \pm 0.3$ & $0.8 \pm 0.2$ & $0.5 \pm 0.2$ & $6.2 \pm 2.3$ & $10.7 \pm 5.0$ & $16.6 \pm 10.5$\\
  \hline
  VCC~50             & $-16.22 \pm 0.09$ & $<14.6$ & $<3.2     \pm 1.9$   & $2.7   \pm 0.3$ & $1.6 \pm 0.2$ & $1.0 \pm 0.2$ & $<1.2 \pm 0.7$ & $<2.0 \pm 1.3$ & $<3.2 \pm 2.1$\\
  VCC~158           & $-16.50 \pm 0.09$ & $<18.6$ & $<8.7     \pm 3.7$   & $3.5   \pm 0.4$ & $2.0 \pm 0.3$ & $1.3 \pm 0.2$ & $<2.5 \pm 1.1$ & $<4.3 \pm 2.0$ &$ <6.7 \pm 3.5$\\
  VCC~538           & $-16.91 \pm 0.09$  & 33.7 & $9.3     \pm 2.3$   & $5.1   \pm 0.6$ & $2.9 \pm 0.4$ & $1.9 \pm 0.4$ & $1.8 \pm 0.5$ & $3.1 \pm 1.0$ & $4.9 \pm 2.0$\\
  VCC~1151 & $-15.80\pm 0.10$ &  \ldots &  \ldots & $1.8\pm0.1$ & $1.1\pm0.2$  & $0.7\pm0.1$  &  \ldots &  \ldots &  \ldots\\  
VCC~1199         & $-16.15  \pm 0.20$ & 58.4 & $13.9   \pm 2.7$   & $2.5   \pm 0.5$ & $1.5 \pm 0.3$ & $0.9 \pm 0.3$ & $5.5 \pm 1.5$ & $9.5 \pm 3.4$ & $14.8 \pm 7.6$\\
  VCC~1440         & $-17.48 \pm 0.09$ & 39.6 & $12.8   \pm 3.4$   & $8.6   \pm 1.0$ & $5.0 \pm 0.7$ & $3.2 \pm 0.6$ & $1.5 \pm 0.4$ & $2.6 \pm 0.8$ &$ 4.0 \pm 1.6$\\
  VCC~1627         & $-17.05 \pm 0.22$ & 49.2 & $10.2   \pm 2.2$   & $5.7   \pm 1.1$ & $3.3 \pm 0.7$ & $2.2 \pm 0.6$ & $1.8 \pm 0.5$ & $3.1 \pm 1.2$ & $4.8 \pm 2.6$\\
  VCC~1896         & $-17.29 \pm 0.09$ & 24.6 & $10.6   \pm 4.2$   & $7.2   \pm 0.8$ & $4.2 \pm 0.6$ & $2.7 \pm 0.5$ & $1.5 \pm 0.6$ & $2.5 \pm 1.1$ & $3.9 \pm 2.0$\\
  VCC~1993         & $-16.82 \pm 0.08$ & 25.1 & $8.2     \pm 3.0$   & $4.7   \pm0.5$ & $2.7 \pm 0.3$ & $1.7 \pm 0.3$ & $1.8 \pm 0.7$ & $3.0 \pm 1.2$ & $4.7 \pm 2.2$\\
    \hline
  CGW~38 & $-17.21\pm0.04$ & 36.4 & $10.63\pm2.5$ & $6.7\pm0.8$ & $3.9\pm0.3$ & $2.5\pm0.2$  & $1.6\pm0.4$ & $2.7\pm0.7$ & $4.3\pm1.1$\\
  CGW~39 & $-16.92\pm0.04$ & 25.7 & $9.5 \pm 4.5$& $5.1\pm0.9$ & $3.0\pm0.3$ & $1.9\pm0.1$ & $1.9\pm0.6$ & $3.2\pm1.0$ & $5.0\pm1.5$\\
  SA426\_002 & $-17.43\pm0.03$ & 33.4 & $35.4\pm11.5$ & $8.2\pm0.6$ & $4.8\pm0.3$ & $3.1\pm0.2$ & $4.3\pm1.1$ & $7.4\pm1.8$ & $11.5\pm2.8$\\
\hline
\end{tabular}
\end{table*}

\subsection{Stellar masses}

Stellar masses are derived for the dEs from their $R$-band magnitudes.   
As these systems are largely devoid of gas, the stellar mass is assumed to be equivalent to the baryonic mass. 
The absolute $R$-band magnitude of each dwarf was converted to a luminosity, and  then multiplied by an $R$-band stellar mass-to-light ($M_{\star}/L_{\star}$) ratio.
We assume that the uncertainty in the distance to the galaxy dominates the error in its luminosity. 
The stellar mass-to-light ratio exhibits variation depending on the simple stellar population (SSP) model and initial mass function (IMF) used, and these variations are explored in \citep{2008MNRAS.386..864D}. 
 In this work, we chose to take our stellar mass-to-light ratios from the SSP models of \citet{2005MNRAS.362..799M}, assuming a Kroupa IMF and a blue horizontal branch. 
However, without an analysis of their stellar populations, age, and metallicity, determining the correct stellar mass-to-light ratio for these galaxies requires additional assumptions. 
 
Dwarf ellipticals exhibit a range of metallicity, and this metallicity can strongly affect their stellar mass-to-light ratios at optical wavelengths. 
For example, dEs in Virgo have metallicity ranging from [Fe/H]$=-1.0$~dex through to [Fe/H]$=+0.03$~dex \citep{2009MNRAS.394.1229C}. 
For a given stellar population of age 10~Gyr, this would result in $R$-band stellar mass-to-light ratios from 2.1 for the lowest metallicity objects, through to 3.3 for the most metal rich dwarfs. 
For the purpose of this work, we assume a constant metallicity when calculating the stellar mass-to-light ratios of these dEs, similar to \citet{2014ApJS..215...17T}. 
We also assume the majority of the dwarfs are metal poor, with  $[Z/H] \approx -0.33$, giving an $R$-band stellar mass-to-light ratio $\sim2.4$.

The stellar mass-to-light ratio increases as the stellar population ages, as low-mass stars contribute less light per unit mass than high-mass stars. 
For example, some dEs in  Virgo reveal evidence for intermediate-age stars \citep[e.g.][]{2012A&A...548A..78T,2014ApJS..215...17T}, with the mean age of their stellar population $\sim{3}$~Gyr. 
If the mean age of the stars were 3~Gyr 
then the $M_{\star}/L_{\star}$ ratio would be systematically lower by a factor of about 2.5 than that of a stellar population with a mean age of 10~Gyr. 
In  \citet{2014ApJS..215...17T}, their figure 16 shows that the majority of discy dEs have derived ages $\sim5$~Gyr.
Furthermore, those exhibiting stellar discs (consistent with an infall and morphological transformation origin) having younger ages than those without discs. 
Likewise, dEs in the group environments may have on average younger stellar population ages than those in a galaxy cluster.
We therefore calculate stellar mass-to-light ratios for three different ages ranging from 3 to 10 Gyr. 
This corresponds to a change of the mass-to-light ratio from 0.9 to 2.4 for our assumed metallicity of $[Z/H]=-0.33$. 

Assuming old stellar populations with age 10~Gyr, the Virgo dwarfs in our sample have half-light stellar masses $1.8\times10^{8} \textrm{~M$_{\odot}$} < M_{\star} < 8.6\times10^{8}$~M$_{\odot}$ i.e. stellar masses within 1~$R_{e}$ (or half of the total stellar mass). 
The majority of group dwarfs have $M_{\star} <10^{9}$~M$_{\odot}$, with the exception of NGC~1407\_13, which has $M_{\star} = 7.7\times10^{9}$~M$_{\odot}$. 

As an additional check, we compared our stellar masses to those obtained from $K$-band photometry.
The $K$-band has the advantage  that it is a good proxy for stellar mass, with the M/L ratio being less sensitive to metallicity variations than optical bands. 
Seven of our galaxies (NGC~1407\_13, VCC~538, VCC~1199, VCC~1440, VCC~1627, VCC~1896, and VCC~1993) are present in the 2MASS extended source catalogue.
For five of the dwarfs (NGC~1407\_13, VCC~538, VCC~1199, VCC~1440 and VCC~1627), their $K$-band stellar masses generally agree to within 50\% of their values derived from $R$-band photometry assuming the same stellar population of age 10~Gyr. 
The two exceptions are VCC~1993 and VCC~1826, which have $R$-band stellar masses more than 50~per~cent higher than their $K$-band stellar masses within 1~$R_{e}$. 
However, these two dwarfs are diffuse, and flux in their outer regions may therefore be missed in their $K$-band photometry.

\subsection{Dynamical masses}
\label{sec:mdyn}

To calculate dynamical masses for our dwarfs, we used the method of \citet{2010MNRAS.406.1220W} who presented a simple formula to calculate  dynamical mass that is largely independent of orbital anisotropy for non-rotating, pressure-supported systems out to one half-light radius $R_{e}$:

\begin{equation}
M_{\rm dyn} = C \mathrm{G^{-1}} \sigma^2 R_e,
\end{equation}

\noindent where  $\sigma$ is a measure of the system's velocity dispersion.

For the size of each system we use 
the  half-light radii $R_e$ listed in Table~\ref{table:size}. 
In principle, the observed $\sigma$ can be corrected  to a uniform standard (such as the
total, luminosity-weighted, infinite-aperture velocity
dispersion) via the variable term $C$ (for further details see discussion by Forbes et al. 2011). 
Here we use the central value of $\sigma$ (see Table 5) calculated at $R_{e}/8$. Using R$_e$ and $\sigma_0$, 
$C$ is well approximated by 6.5 for S\'ersic $n=2$, typical for the dEs examined here. 

Luminous dE galaxies may reveal some rotation \citep{2003AJ....126.1794G,2011A&A...526A.114T}, for which the dynamical masses would need correcting. 
However, as noted above, we find no evidence for rotation in our five low-luminosity Virgo dEs with extended kinematics. 
Therefore we did not correct our values of $\sigma_{0}$ for the effects of rotation prior to calculating the dynamical masses of the dEs.  

To compare the dynamical mass (within the de-projected half-light radius) with the stellar mass, we follow Forbes et al. (2011) and use half of the total stellar mass to calculate the dynamical-to-stellar mass ratio. Half-light values for both stellar mass and dynamical mass are presented in Table~\ref{table:masses} for all objects examined in this study, including the three Perseus dEs from \citet{2014MNRAS.443.3381P}.
For VCC~50, VCC~158, and NGC~1407\_47, the velocity dispersions are likely over-estimated (see Section~\ref{sec:sn}), and therefore their dynamical masses presented in Table~\ref{table:masses} are marked as upper limits.  

\section{Discussion}
\label{sec:discuss}

\subsection{Scaling relations}
\label{sec:sizemag}

\begin{figure*}
\includegraphics[width=0.48\textwidth,height=0.48\textwidth]{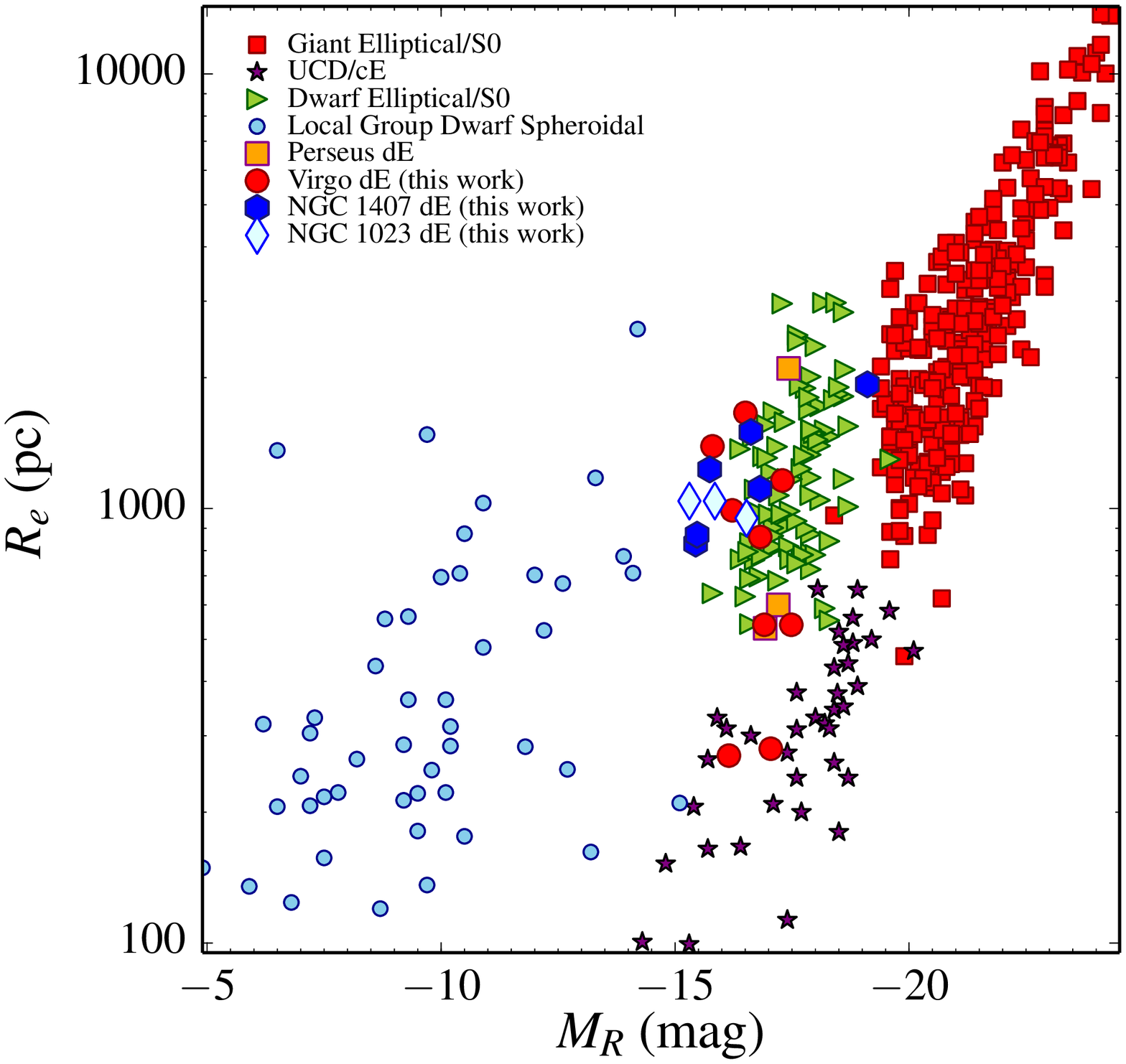}\includegraphics[width=0.48\textwidth,height=0.48\textwidth]{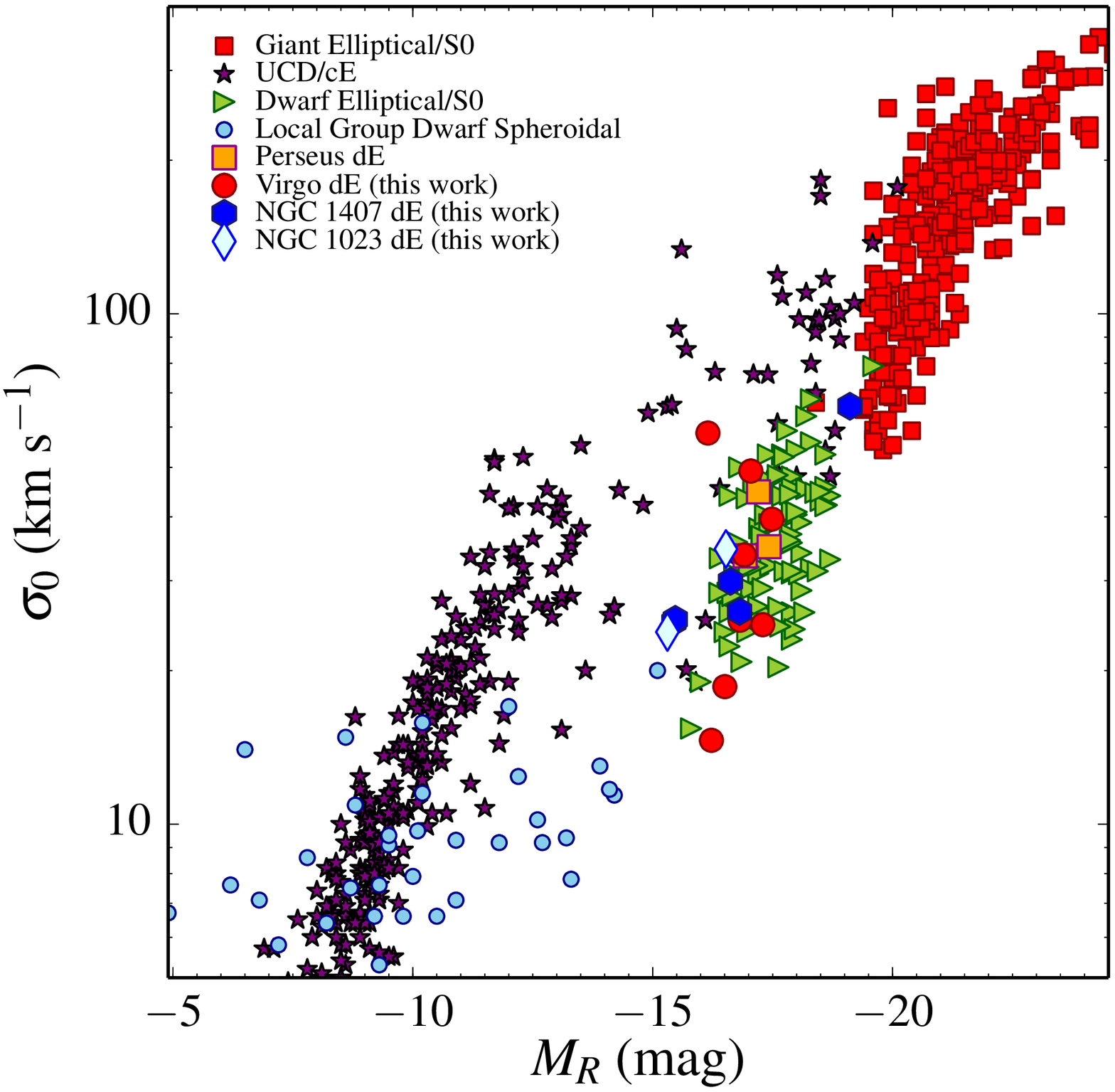}
\caption{Scaling relations for stellar systems across a range of environment. {\it Left panel:} The size-luminosity relation for dEs and other systems in different environments. {\it Right panel:} The sigma-luminosity relation for dEs and other systems. Also included for comparison in both panels is the sample of \citet{2014MNRAS.443.1151N}.  }\label{scaling}
\end{figure*}

We examine the size-magnitude and $\sigma$-luminosity relations for our sample in Figure~\ref{scaling}. 
For comparison to our galaxies, we utilise the sample of \citet{2014MNRAS.443.1151N} and \citet{2014MNRAS.444.2993F}, which provides absolute magnitudes, sizes, velocity dispersions, stellar masses and dynamical masses for compact objects with spectroscopically confirmed distances for objects identified from \textit{Hubble Space Telescope} imaging. 
The sample also includes data for globular clusters, UCDs, dSphs, dEs, cEs, and Es. 
Furthermore, we compliment the comparison sample by data  from \citet{2011MNRAS.414.3699M}, \citet{2011AJ....142..199B} and  \citet{2013MNRAS.435L...6F}, with objects spanning $10^{4}$~M$_{\odot}$ to $10^{12}$~M$_{\odot}$ in stellar mass (globular clusters through to giant ellipticals), and
additional  dEs in the Virgo Cluster from \citet{2014ApJS..215...17T}.

\subsubsection{The size-magnitude relation}
\label{sec:sizelum}

We present the size-magnitude relation for dwarfs in a range of environments in the left-hand panel of Figure~\ref{scaling}. 
The majority of dwarfs targeted in this work have magnitudes fainter than $M_{R} = -18$, the region of the size-magnitude relation where the sizes of dEs/dSphs begin to diverge from those of giant and compact ellipticals.
Regardless of environment, the dEs we examine here follow the same size-magnitude relation in Figure~\ref{scaling}, and the majority overlap in size with dEs taken from the comparison sample of \citet{2014MNRAS.443.1151N}.
 The fainter objects with $M_{R} > -16$~mag  begin to fill in the size-magnitude gap between dSphs and dEs. 
The dEs in the NGC~1023 and NGC~1407 groups have typical sizes for their luminosity, comparable to the brightest dwarfs in the Local Group and the faintest dEs we target in the Virgo Cluster. 
All the group dwarfs we examine here are more extended than cEs of comparable luminosity by $>400$~pc, while the Virgo Cluster and Perseus Cluster objects exhibit a far wider range of size. 

Two of the Virgo Cluster targets, VCC~1199 and VCC~1627, have sizes $R_{e} = 270$~pc and $280$~pc respectively, placing them well below the size-luminosity relation traced  by the majority of dEs at comparable magnitudes ($M_{R} \approx -17$). 
 Instead, in the left-hand panel of Figure~\ref{scaling} they occupy the region of the size-magnitude relation traced by cE galaxies, with sizes comparable to  M32 ($M_{R} = -16$~mag, $R_{e} =113$~pc).
These M32-like objects may be the result of tidal stripping, and we will therefore examine their location on the $\sigma$-luminosity relation in Section~\ref{sec:siglum} to see whether their dynamics are unusual for the luminosity.

\subsubsection{The $\sigma$-luminosity relation}
\label{sec:siglum}

Prior to this work, \citet{2011MNRAS.413.2665F} examined the velocity dispersions of dEs down to $M_{R} \approx -15.7$, and here we extend the study of dE kinematics down to $M_{R} \approx -15.3$ for dEs in groups.  
As can be seen in the right-hand panel of Figure~\ref{scaling}, all dEs in this study follow the same $\sigma$-luminosity relation as objects taken from the literature.
Furthermore, three of the group dEs, NGC~1023\_18, NGC~1407\_47, and NGC~1407\_48, are among the faintest dEs examined to date outside of the Local Group, bridging the luminosity gap between cluster dEs and Local Group dSphs. 
They have comparable values of $\sigma_{0}$ to other faint dEs and the Local Group member NGC~205, and  have lower values of $\sigma_{0}$ than cEs at the same luminosity.  

The $\sigma$-luminosity relation is a useful diagnostic for identifying galaxies with elevated velocity dispersions that may indicate a tidal stripping event. 
When a galaxy is tidally stripped, its luminosity and size will decrease, but its central velocity dispersion will remain relatively unchanged \citep{1992ApJ...399..462B}, decreasing by only a few per~cent \citep[e.g.][]{2009Sci...326.1379C}. 

Two Virgo dEs in our sample, VCC~1199 and VCC~1627, have velocity dispersions $\sigma_{0} = 58.4$~km~s$^{-1}$ and $49.2$~km~s$^{-1}$ respectively, among the highest measured for dEs of their luminosity. 
The high velocity dispersion of VCC~1199 lies on the region of the $\sigma$-luminosity relation occupied by cEs, while the $\sigma_{0}$ value for VCC~1627 is too low to place it on the $\sigma$-luminosity relation seen for cEs of comparable luminosity ($M_{R} = -17.05$). 
As  discussed in Section~\ref{sec:sizelum}, these two galaxies are compact for their luminosity, with sizes $R_{e} < 300$~pc, similar to cEs.
With further tidal stripping, VCC~1627 will reduce in luminosity, and move onto the $\sigma$-luminosity relation for cEs.
To  place it onto the relation, VCC~1627 will need to dim by $\sim1.5$~mag, losing $\approx3\times10^{8}$~M$_{\sun}$, losing 75 per cent of its current stellar mass.

\subsubsection{Dynamical vs. Stellar mass}

While the Local Group dSphs are known to have high dynamical-to-stellar mass ratios, this is unknown for cluster dwarfs due to the difficulty in obtaining deep spectroscopy of faint dwarfs with sufficient resolution to obtain their central velocity dispersions. 
Using stability arguments, Penny et al. (2009) predict that dEs in the Perseus Cluster with stellar masses $<10^{9}$~M$_{\odot}$ are dark matter dominated throughout their structures in order to prevent their disruption by the cluster tidal potential. 
This would result in high mass-to-light ratios in their central regions, and we investigate whether this is the case for the dEs examined in this study by examining how the dynamical masses of the dEs compare to their stellar masses.  

A large dynamical-to-stellar mass ratio may also indicate that a dE formed via tidal stripping. 
Dynamical-to-stellar masses much greater than unity are indicative of either a dark matter dominated object, else a kinematically disturbed or tidally stripped object that is no longer in dynamical equilibrium.

\begin{figure}
\includegraphics[width=0.48\textwidth]{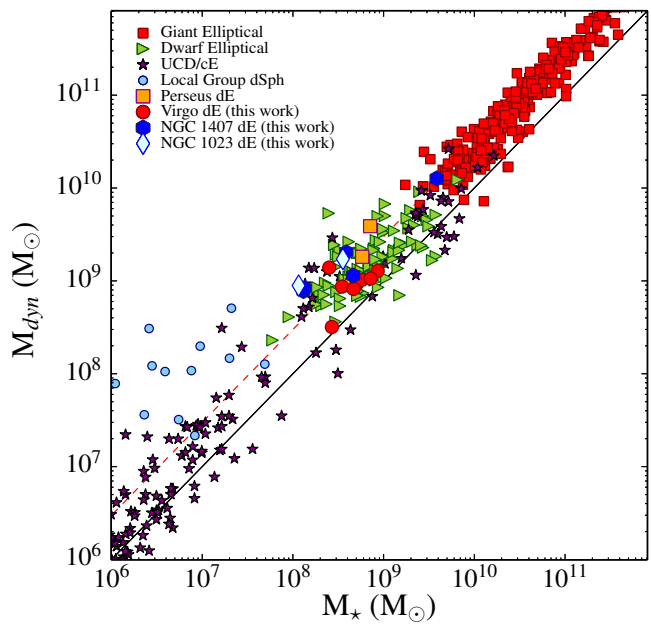}
\caption{Dynamical vs. stellar mass within 1~$R_{e}$ for early-type galaxies and massive star clusters.  
The black line is a 1:1 relation between the dynamical mass $M_{dyn}$, and the stellar mass $M_{\star}$.  The dashed line is a 3:1 ratio. Galaxies that lie above this dashed line have high dynamical-to-stellar mass ratios, similar to the Local Group dSphs. 
It can be seen that the mass ratios of dEs scatter between the 1:1 and 3:1 relations. 
For stellar masses $< 5\times10^{8}$~M$_{\odot}$, three group dEs have elevated dynamical masses, similar to the Local Group dSphs.}\label{dyn_star}
\end{figure}

The dynamical vs. stellar mass diagram for all targeted dwarfs is shown in Figure~\ref{dyn_star}, along with a comparison sample of objects from \citet{2014MNRAS.443.1151N}, with  additional dEs in Virgo provided by \citet{2014ApJS..215...17T}. 
For our sample, the stellar masses within $1$~$R_{e}$ assume an old stellar population with a mean age 10~Gyr and $[Z/H]=-0.33$ as described in Section~\ref{sec:stellar}.
For the comparison sample, we restrict the mass range to objects with masses $10^{6}$~M$_{\odot}$ to $5\times10^{11}$~M$_{\odot}$. 
This restriction in mass was applied to remove globular clusters from the comparison sample, but include UCD-like objects with masses similar to Omega Centauri ($4\times10^{6}$~M$_{\odot}$), hypothesised to be the tidally stripped nucleus of a dE. 
 
We find that Virgo dEs typically have M$_{dyn}$/M$_{\star} < 2$.  
These dwarfs are not dark matter dominated in their central regions, nor do they show evidence for being tidal remnants i.e. having compact sizes for a given luminosity. 

However,  a number of objects in our sample show elevated dynamical-to-stellar mass ratios, regardless of the assumed age of their stellar population. 
Within the Virgo Cluster, VCC~1199 exhibits the most elevated dynamical-to-stellar mass ratios, with $M_{dyn}/M_{\star} = 5.5$ assuming a mean stellar population age of 10~Gyr. 
In Perseus, SA~0426\_002 has the most elevated  dynamical-to-stellar mass ratio, with $M_{dyn}/M_{\star} = 4.2\pm1.1$.
This dwarf is likely tidally interacting with NGC~1275, the central galaxy of the Perseus Cluster \citep{2014MNRAS.443.3381P}, and may not be in dynamical equilibrium as a result of this interaction, resulting in a high value of $\sigma_{0}$ for its luminosity.

The group dEs typically exhibit higher values of $M_{dyn}/M_{\star}$ than those in the Virgo and Perseus clusters, and six of the seven group dEs with measured velocity dispersions have  $M_{dyn}/M_{\star} > 3$.  
Assuming stellar populations of age 10~Gyr, the cluster dEs (Perseus and Virgo) have a mean value $M_{dyn}/M_{\star} = 2.2 \pm 0.5$, vs $M_{dyn}/M_{\star} = 5.1 \pm 0.6$ for the group dEs. 
Lowering the mean stellar age of these objects will increase $M_{dyn}/M_{\star}$, as will decreasing their metallicities.  
However, the group targets extend to lower luminosities than the cluster targets, and it can be seen in Fig.~\ref{dyn_star} that lower luminosity galaxies exhibit higher values of $M_{dyn}/M_{\star}$.
Our sample size, particularly for fainter galaxies, it not large enough to examine luminosity trends in $M_{dyn}/M_{\star}$ with respect to environment.

\subsection{The Virgo Cluster}  

\begin{figure}
\includegraphics[width=0.48\textwidth]{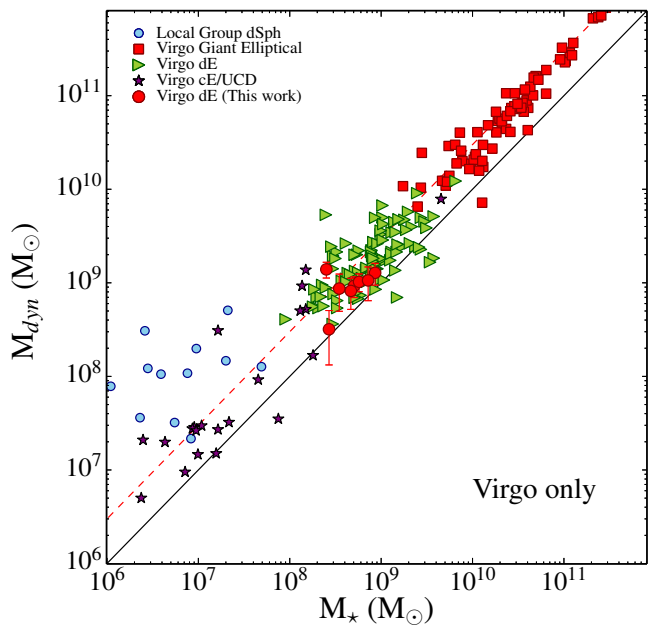}
\caption{Dynamical vs. stellar mass within 1~$R_{e}$ for early-type galaxies and massive star clusters in the Virgo Cluster. 
As in previous plots, dSphs from the Local Group are included for comparison. 
The solid black line is a 1:1 relation between $M_{dyn}$ and $M_{\star}$, and the dashed red line is a 3:1 relation. 
The eight Virgo dEs presented in this work fall within the scatter of previously studied Virgo dEs, and none exhibit elevated values of $M_{dyn}$ for their stellar mass. }\label{virg_mems}
\end{figure}

Due to its close distance of 16.7~Mpc  \citep{2009ApJ...694..556B}, large population of dEs with confirmed cluster membership, and deep imaging from which to obtain sizes, Virgo is the most common cluster in which to study dE kinematics. 
These dEs share a common environment and similar distance, removing some of the uncertainty in their size determination. 
We therefore re-plot Figure~\ref{dyn_star} for Virgo Cluster members only in Figure~\ref{virg_mems}. 
The Local Group dSphs are included for comparison, as no internal kinematics have been determined for early-type dwarfs less massive than $\sim5\times10^{7}$~M$_{\sun}$ in the Virgo Cluster (or indeed, any environment outside the Local Group).

We compare our galaxies to the Virgo Cluster dE samples of Geha et al. (2003), Forbes et al. (2011), Rys et al. (2013), and \citet{2014ApJS..215...17T}. 
Virgo Cluster dEs present in the catalogue of \citet{2014MNRAS.443.1151N} are also included in Figure~\ref{virg_mems}.
Three of the compact Virgo dwarfs in our sample VCC~1199, VCC~1440, and VCC~1627 have also had their dynamical masses determined by \citet{2015ApJ...804...70G} using velocity dispersions obtained from Gemini-GMOS IFU spectroscopy. 
Within the error bars, our dynamical masses agree with those obtained in their study, though we do not include these galaxies in Figure~\ref{virg_mems} for simplicity.  
The majority of our Virgo dE targets follow the $M_{dyn}$ vs. $M_{\star}$ relation traced by Virgo giant ellipticals, though while the majority of giant Es trace a 3:1 relation, the dEs scatter more between a 3:1 and 1:1 relation. 
None of the Virgo dEs/cEs we examine exhibit elevated values of $M_{dyn}$ for their stellar mass. 

\subsubsection{Mass-to-light ratio vs. cluster-centric distance}

\begin{figure}
\includegraphics[width=0.47\textwidth]{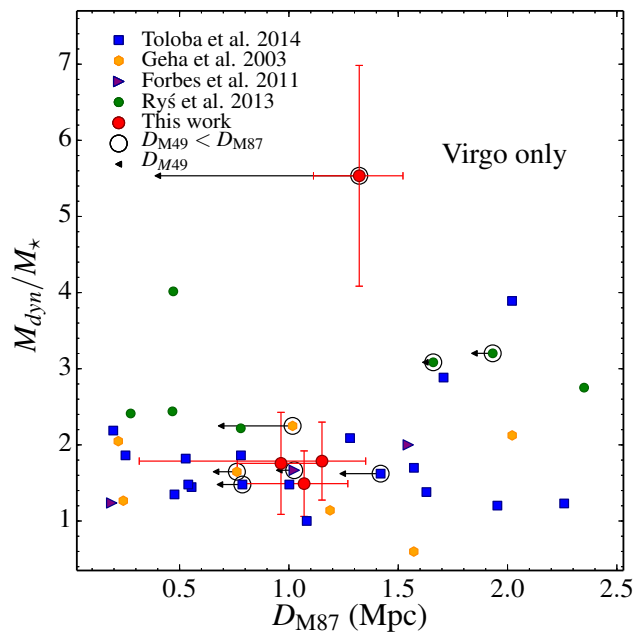}
\caption{Dynamical-to-stellar mass ratio $M_{dyn}/M_{\star}$ vs. ``de-projected'' distance to M87 for dEs in the Virgo Cluster. Also included for comparison are the results of \citet{2003AJ....126.1794G}, \citet{2011MNRAS.413.2665F}, \citet{2013MNRAS.428.2980R}, and \citet{2014ApJS..215...17T}. dEs with a smaller separation from M49 than M87 are circled, and their distances to M49 plotted as triangles. An arrow links the two distances. No obvious  trend is seen between $M_{dyn}/M_{\star}$ and cluster-centric distance.}\label{ratvdis} 
\end{figure}

\citet{2014MNRAS.439..284R} tentatively identified a trend such that the dynamical-to-stellar mass ratio increases as a function of distance from the centre of the Virgo Cluster. 
The dEs with highest values of $M_{dyn}/M_{\star}$ are at distances $>1$~Mpc from M87, which is similar to the virial radius $R_{vir} = 1.08$~Mpc \citep{2005A&A...435....1P} of the Virgo A subcluster. 
This relation was found when  ``de-projecting'' the distances of their Virgo dE sample using redshift independent distances in combination with angular separations from M87, removing some of the uncertainty in the distances of the objects from the cluster-centric galaxy M87. 

We investigate whether a similar relation is seen when enlarging the sample with 
 our Virgo dEs, de-projecting their distances from M87 with redshift independent distances from the literature. 
Distances from surface brightness fluctuations are found for 5 dEs in our sample (VCC~1199, \citealt{2010ApJ...724..657B}; VCC~523, VCC~1440, VCC~1627, and VCC~1993, \citealt{2009ApJ...694..556B}). 
We combine these distances with the angular separations of the dEs to M87 to provide a de-projected distance to M87.

We plot $M_{dyn}/M_{\star}$ vs. de-projected cluster-centric distance in Figure~\ref{ratvdis} for the Virgo Cluster members.
Four of the dEs have distances $<1.2$~Mpc from the cluster centre.
However, VCC~538 is located at a distance of $23.0$~Mpc, 6.3~Mpc more distant than M87 at $16.7$~Mpc. 
Virgo has complex substructure, and this distance places  VCC~538  outside the Virgo A  cluster (around M87).Its angular separation and radial velocity place it in the Virgo B subcluster, at a projected separation of 1~Mpc from M49, though it is at de-projected distance of 6.3~Mpc from M49, which again places it behind the cluster.
VCC~538 is therefore not included  in Figure~\ref{ratvdis}. 
Additionally we include data from \citet{2003AJ....126.1794G}, \citet{2011MNRAS.413.2665F}, \citet{2013MNRAS.428.2980R} and \citet{2014ApJS..215...17T} for which surface brightness fluctuation distances are available.
The majority of the surface brightness fluctuation distances for both samples are taken from \citet{2009ApJ...694..556B}, with the distance to three dEs from \citet {2014ApJS..215...17T} provided by \citet{2004AJ....127..771J}. This adds 7 dEs from \citet{2013MNRAS.428.2980R} and 19 dEs from \citet{2014ApJS..215...17T} to the plot. 

It can be seen from Figure~\ref{ratvdis} that there is no clear trend for dEs at larger cluster-centric distance to exhibit higher values of $M_{dyn}/M_{\star}$ than those at small cluster-centric distance.  
Some dEs in our sample lie closer in distance to M49, the brightest galaxy in the Virgo Cluster, and we highlight such galaxies with small separation from M49 by circling them in Figure~\ref{ratvdis}. 
Despite highlighting galaxies that are at a smaller separation from M49 than M87, no trend is found in $M_{dyn}/M_{\star}$ with cluster-centric distance with the remaining satellites of M87.

While the dEs with higher values of $M_{dyn}/M_{\star}$ at first glance seem to favour large separations from M87, the samples contain dwarfs over 3 orders of magnitude in brightness.
In our Virgo sample, along with both the \citet{2003AJ....126.1794G} and \citet{2014ApJS..215...17T} comparison samples, the fainter dwarfs exhibit higher values of $M_{dyn}/M_{\star}$.
Thus more kinematics of Virgo dEs at low luminosity are required to separate trends of $M_{dyn}/M_{\star}$ with respect to both galaxy luminosity and cluster-centric distance.

\subsubsection{VCC~538, VCC~1199 and VCC~1627- compact ellipticals?}
\label{sec:compacts}

Compact ellipticals typically have radii $100$~pc~$< R_{e} < 500$~pc, and they are hypothesised to be the central remnants of intermediate-mass galaxies \citep[e.g.][]{1973ApJ...179..423F}.
Such galaxies have similar scaling parameters to M32, the prototype compact elliptical (cE) galaxy, having small effective radii for their respective magnitudes, similar to M32 with $R_{e} = 113$~pc and $M_{R} \approx-15.9$.
As such, they are interesting objects in which to study scaling relations and galaxy dynamics, as their central velocity dispersions can be used to help identify the nature of their progenitor \citep[e.g][]{2014MNRAS.439.3808P}. 
During the formation of such objects, tidal interactions are an effective way to slow down or even stop rotation, transforming a rotationally supported galaxy into a pressure supported one \citep[e.g.][]{2001Ap&SS.276..375M,2001ApJ...547L.123M}.

Five of our target dwarfs are classified as E0/E2 galaxies in NED, and two of them,  VCC~1199, and VCC~1627,  have effective radii $R_{e}\sim300$~pc. As such, their sizes and luminosities place them in the region of the size-magnitude relation occupied by cE galaxies (Figure~\ref{scaling}), and these two galaxies may be the central bulges of tidally stripped disc of elliptical galaxies. 

The remaining three galaxies classified as ellipticals have  $R_{e}$ larger than $500$~pc, with $R_{e} = 540$~pc VCC~538, $R_{e} = 540$~pc for VCC~1440, and $R_{e}=860$~pc for VCC~1993, comparable in size to dEs at similar luminosity. 
Furthermore, all five of the dwarfs classified as Es have kinematics resolved beyond their centres, allowing us to determine if they are dominated by rotational or pressure support. 
None of the five exhibit rotation out to $R=300$~pc, suggesting they have undergone sufficient tidal interactions to remove evidence of rotation in their kinematics.
We also note that there is ambiguity in the true size of VCC~538, as its radial velocity places it at a similar distance as M87 ($D=16.7$~Mpc), where it would have $R_{e} = 380$~pc, placing it on the cE sequence in the size-magnitude relation. 
Compact ellipticals are rare objects, and VCC~538, VCC~1199 and VCC~1627 have been claimed to belong to this class of galaxy \citep{1985AJ.....90.1681B}. 
We therefore discuss the nature of VCC~538, VCC~1199, and VCC~1627 as  cEs vs. dEs below.

\smallskip
\emph{VCC~538} lies $\sim$7~Mpc from the Virgo Cluster central galaxy M87, with surface brightness fluctuations placing VCC~538 at $D=23.0$~Mpc.
Though it is closer in projection to the Virgo brightest cluster galaxy M49 than to M87, its surface brightness fluctuation distance places it more distant than M49 by $\sim6$~Mpc. 
At $D=23.0$~Mpc, its effective radius $R_{e} = 4\farcs8$ corresponds to $535$~pc, and it would have $M_R = -16.9$.
While on the low side, its size $R_{e}=535$~pc is not low enough to qualify VCC~538 as a compact elliptical. 
Instead, we classify it as a dE. 
Its value of $M_{dyn}/M_{\star} = 1.8$ is similar to dEs of comparable magnitude in our sample.

\smallskip
\emph{VCC~1199} has the most elevated value of velocity dispersion ($\sigma_{0} =58.4$~km~s$^{-1}$), smallest size ($R_{e} = 270$~pc), and highest value of dynamical-to-stellar mass ratio ($M_{dyn}/M_{\star} = 5.5$) of our Virgo dwarf targets. 
It is furthermore extremely red in colour for a dE of its magnitude, with $(g-z) = 1.56$ \citep{2006ApJS..164..334F}. 
It is offset from the red sequence of Virgo dEs  by $\sim0.5$~mag, with a colour comparable to a typical red sequence galaxy with $M_{g} < -20$~mag. 
This  red colour is typical for cEs of comparable magnitude \citep[e.g.][]{2013ApJ...772...68S}, and is indicative of a tidal stripping event.
As in previous studies, we hypothesise that this object is currently undergoing tidal stripping, having undergone enough tidal interactions to remove evidence of rotation in its radial velocity profile.
It is at a de-projected distance $\sim$0.4~Mpc from M49, and therefore resides in one of the densest parts of the Virgo Cluster where harassment by other cluster members is likely. 
\citet{2008ApJ...681..197P}  stated that VCC~1199 does not have a detectable globular cluster system, again consistent with it being a tidally stripped system.  

In addition, \citet{2006ApJS..164..334F} commented that the galaxy exhibits a spiral pattern, is tidally truncated by M49, and is a prime candidate for galaxy harassment. 
VCC~1199 is therefore likely a dynamically young object, with the spiral structure of its progenitor not yet erased by tidal interactions, else the spiral arms could be tidally induced.
It exhibits a cuspy profile with an excess of light at its centre when fit with a single profile S\'ersic fit \citep{2009ApJS..182..216K}, and this excess light may have been produced in a past tidal interaction. 
The object is offset from $\sigma$-luminosity locus for cEs by less than 0.5~mag and $<10$~km~s$^{-1}$. However, the error on its velocity dispersion is $\pm6.1$~km~s$^{-1}$, enough to move it onto the cE locus, and we therefore consider it to be a cE galaxy.

\smallskip
\emph{VCC~1627} is the second most compact object in our sample ($R_{e} =290$~pc), with a high central velocity dispersion ($\sigma_{0}=49.2\pm6.2$~km~s$^{-1}$). 
Despite its extreme size and velocity dispersion,  assuming an old stellar population, VCC~1627 has a low value of $M_{dyn}/M_{\star} = 1.8 \pm 0.77$, i.e. it does not have an unusual dynamical mass for its luminosity.  
Based on its size alone, VCC~1627 would be considered a cE as it lies on the cE locus of the size-magnitude relation in Figure~\ref{scaling}. 
However, its velocity dispersion places it off the $\sigma$-luminosity locus for cEs by $\sim20$~km~s$^{-1}$, though similar sized galaxies are classified as cEs in Figure~\ref{scaling}.
We therefore tentatively classify VCC~1627 as a cE.  

In their recent study of the dynamical-to-stellar mass ratio of UCDs and cEs, \citet{2014MNRAS.444.2993F} found half a dozen objects to have extremely high mass ratios. 
One of those objects was VCC~1627 with a mass ratio of almost 7. 
The key parameters for VCC~1627 came from the compilation of \citet{2014MNRAS.443.1151N}.
In this case, their velocity dispersion of 47.3~km~s$^{-1}$ is from the Sloan Digital Sky Survey \citep[SDSS,][]{2000AJ....120.1579Y}, and is in good agreement with our central $1\farcs2$ aperture value of $49.2 \pm 1.2$~km~s$^{-1}$. 
Their half-light radius of 274.2~pc compares well with our value $R_{e} =290$~pc. 
The main difference between this work and \citet{2014MNRAS.443.1151N} is in the value used for the stellar mass. 
Using a $V$-band magnitude of $M_V = -16.7$ and a Kroupa IMF, \citet{2014MNRAS.443.1151N}  calculated a total stellar mass of $1.37 \times$ 10$^{8}$~M$_{\odot}$.

Here we used an $R$-band magnitude of $M_R = -17.04$ assuming the dE is at a distance of 15.6~Mpc, which provided a half-light luminosity of $2.7\times10^{8}$~L$_{\odot}$.   
In Table 5 we list the stellar mass converted from this luminosity corresponding to mean stellar ages of 10, 5 and 3~Gyr for a Kroupa IMF and metallicity of $[Z/H]=-0.33$~dex. 
The weak Balmer lines in the \textsc{esi} spectrum of VCC~1627 suggest a very old age and hence a half-light stellar mass of about $6 \times 10^{8}$~M$_{\odot}$ assuming a mean stellar age of 10~Gyr. 
This is in close agreement with our stellar mass calculated using the metallicity-insensitive $K$-band, which provides a half-light stellar mass $\sim5.4\times10^{8}$~M$_{\odot}$.
Regardless of the age of the stellar population, the three stellar masses we calculated are all greater than that of \citet{2014MNRAS.443.1151N}. 
However, if we take distance of  VCC~1627 to be 7.7~Mpc, based on its velocity of 236~km~s$^{-1}$ as provided by the NASA Extragalactic Database (NED), then we re-calculate a stellar mass of $1.37 \times$ 10$^{8}$~M$_{\odot}$, identical to that used in  \citet{2014MNRAS.444.2993F} and \citet{2014MNRAS.443.1151N}.
 The distance to VCC~1627 provided by NED would place the dwarf well outside the Virgo Cluster, thus we use its distance from surface brightness fluctuations. 
We therefore obtain a dynamical-to-stellar mass ratio of $\sim1.8$ assuming a mean stellar age of 10~Gyr and a distance of 15.6~Mpc. 
Our value of $M_{dyn}/M_{\star}$ is no longer an extreme mass ratio, and is only slightly elevated above unity.

\subsubsection{Flat velocity dispersion and radial velocity profiles}

The majority of the Virgo dEs in our sample with extended kinematic measurements have nearly flat velocity dispersion profiles, with only VCC~1199 exhibiting a clear central peak. 
Flat velocity dispersion profiles  in dEs are also found by \citet{2011MNRAS.413.2665F}, \citet{2011A&A...526A.114T}, and \citet{2013MNRAS.428.2980R}. 
A comparable result is observed for the Local Group dSphs, which exhibit flat velocity dispersion profiles out to $\sim1$~kpc from their centres (e.g.\ Walker et al. 2007).
Assuming these galaxies are in virial equilibrium, this is taken as evidence that the Local Group dSphs possess extended dark matter haloes that dominate their total mass. 
However, the dEs typically have values of $M_{dyn}/M_{\star} < 2$ within 1~$R_{e}$, implying their central kinematics are not dark matter dominated, comparable to the results of \citet{2011A&A...526A.114T}. 
The lack of central peaks in the velocity dispersion profiles of the dwarfs also show that their central kinematics are not dominated by intermediate-mass black holes.

Furthermore, we do not detect any rotation in the same five Virgo dwarfs to within $\pm10$~km~s$^{-1}$ of their central values, with typical errors on their measured radial velocities $\pm6$~km~s$^{-1}$. 
All five dwarfs have cluster-centric distance that place them within one virial radius of the Virgo A / M87 subcluster, though VCC~538 has a distance from surface brightness fluctuations that places it outside the main cluster. 
VCC~1199 is furthermore likely in the Virgo B / M49 subcluster, again subject to a high local galaxy density favourable to numerous galaxy-galaxy interactions. 
With no evidence of rotation in their kinematics, the five dEs in our sample with extended kinematics are slow rotators. 
Given their central location and lack of rotation, these objects have undergone sufficient tidal interactions that all evidence for rotation has been erased from their kinematics \citep[see the simulations of e.g.][]{2005MNRAS.364..607M}. 

Toloba et al.\ (2014) find a similar correlation for dEs in Virgo, such that 10 of the 11 slow rotators in their sample are found within 1~Mpc of the Virgo Cluster central galaxy M87. 
After several passes through the cluster centre, the progenitor is heated, and transformed into a pressure supported system, and objects at smaller cluster-centric radii will be transformed more quickly. 
Therefore, if the dEs and cEs in our sample have undergone enough interactions, they will not exhibit rotation in their kinematics, and are pressure supported systems.  

\subsection{Group dEs}

We present velocity dispersions for seven dEs in the NGC~1023 and NGC~1407 groups. 
The two groups are very different: NGC~1023 is an un-evolved group, consisting of a few spiral galaxies, whereas NGC~1407 is often taken to be a nearby analogue of a fossil group, dominated by its central elliptical. 
Despite their very different environments to the Virgo and Perseus clusters, the group dEs follow the same size-magnitude and $\sigma$-luminosity relations as objects in higher density environments. 
For both groups, the group dEs with $M_{R} > -16$ lie on the branch in the size-magnitude relation traced by local group dSphs and the brighter dE galaxies. 

The NGC~1023 group is remarkably lacking in bright, passive dEs, with a dE fraction 45~percent of all dwarfs, vs. a dE fraction of 80~percent of all dwarfs in the NGC~1407 group \citep{2009MNRAS.398..722T}. 
Indeed, two of the dwarfs we investigate, NGC~1023\_14 and NGC~1023\_18, have ongoing star formation. NGC~1023\_18 has a large separation $>1$~Mpc from NGC~1023, and is thus unlikely to have been quenched by its group environment. 
However, NGC~1023\_18 is located at a projected distance $\sim120$~kpc from NGC~1023, where it might be expected to be tidally stripped by its host galaxy, and may therefore be tidally perturbed, increasing its velocity dispersion. 
If bright dEs are the remnants of tidally stripped disc galaxies, sparse groups such as the Local Group and the NGC~1023 group are likely inefficient at transforming infalling discs into objects with little or no star formation. 

Six of the seven group dEs examined here have elevated values of $M_{dyn}/M_{\star} > 3$, with the fainter dEs with $M_{R} > -15.5$ exhibiting the highest values. 
The group dEs fall into the luminosity gap where few dEs outside of the Local Group have measured kinematics. 
From the left-hand panel of Figure~6, these dEs in the NGC~1023 and NGC~1407 groups exhibit  values of $M_{dyn}/M_{\star} > 3$, comparable to those of the brightest dSphs in the Local Group. 
These elevated values of $M_{dyn}/M_{\star} > 3$ suggest that like the Local Group dSphs, the faintest dEs in our sample are dominated by dark matter within 1~$R_{e}$. 
Our calculated values of $M_{dyn}/M_{\star}$ for the faint dEs also agree with those predicted in \citet{2009MNRAS.393.1054P} for dEs of comparable luminosity in the Perseus Cluster.  
However, we caution that the faintest dEs in this study have velocity dispersions comparable to the resolution of Keck-\textsc{esi}, and their true velocity dispersions, and hence dynamical masses, may be lower than those measured here.

Dwarf ellipticals in low density environments typically have younger mean stellar ages than those in cluster cores.
 For example, within the Coma Cluster, dEs at a cluster-centric distance of 2.5~Mpc have mean stellar ages $\sim$half that of dEs in the centre of Coma \citep{2012MNRAS.419.3167S}. 
Thus we may be over-estimating the $R$-band stellar mass-to-light ratios for the group dEs by assuming they have the same mean stellar ages as those in the Virgo Cluster. 
This effect is clearly seen in Table~\ref{table:masses}, where the stellar masses calculated using $R$-band luminosities are lower for a younger stellar population.  
Ideally, we would determine the stellar masses for all the dEs in our sample using $K$-band photometry, for which the stellar mass-to-light ratios are fairly insensitive to stellar age for galaxies with mean stellar ages $>3$~Gyr.  
Unfortunately, well-measured $K$-band magnitudes are not available for the majority of our group dEs, as they are too low surface brightness to be detected in the 2MASS survey. Therefore the calculated values assuming a mean stellar age of 10~Gyr are upper limits on their stellar mass, and the true values of $M_{dyn}/M_{\star}$ for these objects are likely higher.

\section{Conclusions}
\label{sec:conclude}

In this work, we have measured velocity dispersions for dE galaxies in three environments: the Virgo Cluster, the E/S0 dominated NGC~1407 group, and the spiral dominated NGC~1023 group. 
We supplemented these data with three dEs in the Perseus Cluster first presented in \citet{2014MNRAS.443.3381P}. 
While the majority of studies to date have focussed on dEs in the cluster environment, we have extended the study of these objects to less dense group regions. 
Central velocity dispersions were measured for 15 dwarf using the Keck-\textsc{esi} spectrograph, with velocity dispersion profiles measured for five compact dEs/cEs in the Virgo Cluster. 

We found that the five compact dEs in the Virgo Cluster with sizes $ {R_{e}} < 500$~pc have flat velocity dispersion profiles out to 300~pc from the centres ($\sim1$~$R_{e}$ for the three most compact objects in our sample), with no evidence for central peaks as found for more massive galaxies such as giant ellipticals.
We furthermore found no evidence for rotation in these 5 compact Virgo dwarfs to the same radius, and they are instead pressure supported slow rotators. 
Two of the Virgo galaxies, VCC~1199 and VCC~1627 have sizes $R_{e} <300$~pc, and velocity dispersions $\sim50$~km~s$^{-1}$, and we therefore classify these objects as compact ellipticals. 

Despite the range of environments examined here, from the spiral dominated NGC~1023 group, the E/S0 dominated NGC~1407 group, through to the Virgo and Perseus clusters, the dEs and cEs we examined here follow the same $\sigma$-luminosity and size-magnitude relations. 
There is a slight preference for dEs in groups to exhibit higher values of $M_{dyn}/M_{\star}$, with a mean value of $M_{dyn}/M_{\star}=5.1\pm0.6$ for group dEs, vs. $M_{dyn}/M_{\star}=2.2\pm0.5$ for cluster dEs. 
We note that this difference may be due to the group dwarfs having lower luminosity than our Virgo Cluster targets, else they could host different stellar populations.
We also search for trends in $M_{dyn}/M_{\star}$ vs. distance from M87 for dEs in the Virgo Cluster, by ``de-projecting'' their distances from M87 using surface brightness fluctuation distances. 
We supplemented our dE sample of five Virgo dwarfs with redshift independent distances with data taken from the literature.  
No trend is found between $M_{dyn}/M_{\star}$ and distance from M87, with no preference for dEs with high values of  $M_{dyn}/M_{\star}$ to reside in the outskirts or centre of the Virgo Cluster.

\section*{Acknowledgements}
SJP acknowledges the support of an Australian Research Council Super Science Postdoctoral Fellowship grant FS110200047 and postdoctoral funding from the University of Portsmouth. 
JJ and DAF thank the ARC for financial support via DP130100388.
We thank Caroline Foster, Nicola Pastorello, and Vincenzo Pota for their assistance with our observations.  
We thank the referee whose comments helped improve this paper.
This project made use of the
NASA Extragalactic Database (NED) and 
data products from the Two Micron All Sky Survey, which is a
joint project of the University of Massachusetts and the Infrared
Processing and Analysis Center/California Institute of
Technology, funded by the National Aeronautics and Space
Administration and the National Science Foundation. We thank the
staff of the W. M. Keck Observatory for their support. Some the
data presented herein were obtained at the W.M. Keck
Observatory, which is operated as a scientific partnership among the
California Institute of Technology, the University of California and
the National Aeronautics and Space Administration. 
This work was supported in part by National Science Foundation Grant No. PHYS-1066293 and the hospitality of the Aspen Center for Physics.
The authors wish to recognise and acknowledge the very significant cultural role and reverence that the summit of Mauna Kea has always had within the indigenous Hawaiian community.  
We are most fortunate to have the opportunity to conduct observations from this mountain.




\bibliographystyle{mnras}
\bibliography{spenny} 

\begin{thebibliography}{}
\makeatletter
\relax
\def\mn@urlcharsother{\let\do\@makeother \do\$\do\&\do\#\do\^\do\_\do\%\do\~}
\def\mn@doi{\begingroup\mn@urlcharsother \@ifnextchar [ {\mn@doi@}
  {\mn@doi@[]}}
\def\mn@doi@[#1]#2{\def\@tempa{#1}\ifx\@tempa\@empty \href
  {http://dx.doi.org/#2} {doi:#2}\else \href {http://dx.doi.org/#2} {#1}\fi
  \endgroup}
\def\mn@eprint#1#2{\mn@eprint@#1:#2::\@nil}
\def\mn@eprint@arXiv#1{\href {http://arxiv.org/abs/#1} {{\tt arXiv:#1}}}
\def\mn@eprint@dblp#1{\href {http://dblp.uni-trier.de/rec/bibtex/#1.xml}
  {dblp:#1}}
\def\mn@eprint@#1:#2:#3:#4\@nil{\def\@tempa {#1}\def\@tempb {#2}\def\@tempc
  {#3}\ifx \@tempc \@empty \let \@tempc \@tempb \let \@tempb \@tempa \fi \ifx
  \@tempb \@empty \def\@tempb {arXiv}\fi \@ifundefined
  {mn@eprint@\@tempb}{\@tempb:\@tempc}{\expandafter \expandafter \csname
  mn@eprint@\@tempb\endcsname \expandafter{\@tempc}}}

\bibitem[\protect\citeauthoryear{{Bender}, {Burstein}  \& {Faber}}{{Bender}
  et~al.}{1992}]{1992ApJ...399..462B}
{Bender} R.,  {Burstein} D.,   {Faber} S.~M.,  1992, \mn@doi [\apj]
  {10.1086/171940}, \href {http://adsabs.harvard.edu/abs/1992ApJ...399..462B}
  {399, 462}

\bibitem[\protect\citeauthoryear{{Binggeli}, {Sandage}  \&
  {Tammann}}{{Binggeli} et~al.}{1985}]{1985AJ.....90.1681B}
{Binggeli} B.,  {Sandage} A.,   {Tammann} G.~A.,  1985, \mn@doi [\aj]
  {10.1086/113874}, \href {http://adsabs.harvard.edu/abs/1985AJ.....90.1681B}
  {90, 1681}

\bibitem[\protect\citeauthoryear{{Blakeslee} et~al.,}{{Blakeslee}
  et~al.}{2009}]{2009ApJ...694..556B}
{Blakeslee} J.~P.,  et~al., 2009, \mn@doi [\apj] {10.1088/0004-637X/694/1/556},
  \href {http://adsabs.harvard.edu/abs/2009ApJ...694..556B} {694, 556}

\bibitem[\protect\citeauthoryear{{Blakeslee} et~al.,}{{Blakeslee}
  et~al.}{2010}]{2010ApJ...724..657B}
{Blakeslee} J.~P.,  et~al., 2010, \mn@doi [\apj] {10.1088/0004-637X/724/1/657},
  \href {http://adsabs.harvard.edu/abs/2010ApJ...724..657B} {724, 657}

\bibitem[\protect\citeauthoryear{{Brodie}, {Romanowsky}, {Strader}  \&
  {Forbes}}{{Brodie} et~al.}{2011}]{2011AJ....142..199B}
{Brodie} J.~P.,  {Romanowsky} A.~J.,  {Strader} J.,   {Forbes} D.~A.,  2011,
  \mn@doi [\aj] {10.1088/0004-6256/142/6/199}, \href
  {http://adsabs.harvard.edu/abs/2011AJ....142..199B} {142, 199}

\bibitem[\protect\citeauthoryear{{Cappellari} et~al.,}{{Cappellari}
  et~al.}{2006}]{2006MNRAS.366.1126C}
{Cappellari} M.,  et~al., 2006, \mn@doi [\mnras]
  {10.1111/j.1365-2966.2005.09981.x}, \href
  {http://adsabs.harvard.edu/abs/2006MNRAS.366.1126C} {366, 1126}

\bibitem[\protect\citeauthoryear{{Chen}, {C{\^o}t{\'e}}, {West}, {Peng}  \&
  {Ferrarese}}{{Chen} et~al.}{2010}]{2010ApJS..191....1C}
{Chen} C.-W.,  {C{\^o}t{\'e}} P.,  {West} A.~A.,  {Peng} E.~W.,   {Ferrarese}
  L.,  2010, \mn@doi [\apjs] {10.1088/0067-0049/191/1/1}, \href
  {http://adsabs.harvard.edu/abs/2010ApJS..191....1C} {191, 1}

\bibitem[\protect\citeauthoryear{{Chilingarian}}{{Chilingarian}}{2009}]{2009MNRAS.394.1229C}
{Chilingarian} I.~V.,  2009, \mn@doi [\mnras]
  {10.1111/j.1365-2966.2009.14450.x}, \href
  {http://adsabs.harvard.edu/abs/2009MNRAS.394.1229C} {394, 1229}

\bibitem[\protect\citeauthoryear{{Chilingarian}, {Cayatte}, {Revaz}, {Dodonov},
  {Durand}, {Durret}, {Micol}  \& {Slezak}}{{Chilingarian}
  et~al.}{2009}]{2009Sci...326.1379C}
{Chilingarian} I.,  {Cayatte} V.,  {Revaz} Y.,  {Dodonov} S.,  {Durand} D.,
  {Durret} F.,  {Micol} A.,   {Slezak} E.,  2009, \mn@doi [Science]
  {10.1126/science.1175930}, \href
  {http://adsabs.harvard.edu/abs/2009Sci...326.1379C} {326, 1379}

\bibitem[\protect\citeauthoryear{{C{\^o}t{\'e}} et~al.,}{{C{\^o}t{\'e}}
  et~al.}{2004}]{2004ApJS..153..223C}
{C{\^o}t{\'e}} P.,  et~al., 2004, \mn@doi [\apjs] {10.1086/421490}, \href
  {http://adsabs.harvard.edu/abs/2004ApJS..153..223C} {153, 223}

\bibitem[\protect\citeauthoryear{{D'Onghia}, {Besla}, {Cox}  \&
  {Hernquist}}{{D'Onghia} et~al.}{2009}]{2009Natur.460..605D}
{D'Onghia} E.,  {Besla} G.,  {Cox} T.~J.,   {Hernquist} L.,  2009, \mn@doi
  [\nat] {10.1038/nature08215}, \href
  {http://adsabs.harvard.edu/abs/2009Natur.460..605D} {460, 605}

\bibitem[\protect\citeauthoryear{{Dabringhausen}, {Hilker}  \&
  {Kroupa}}{{Dabringhausen} et~al.}{2008}]{2008MNRAS.386..864D}
{Dabringhausen} J.,  {Hilker} M.,   {Kroupa} P.,  2008, \mn@doi [\mnras]
  {10.1111/j.1365-2966.2008.13065.x}, \href
  {http://adsabs.harvard.edu/abs/2008MNRAS.386..864D} {386, 864}

\bibitem[\protect\citeauthoryear{{De Rijcke}, {Dejonghe}, {Zeilinger}  \&
  {Hau}}{{De Rijcke} et~al.}{2003}]{2003A&A...400..119D}
{De Rijcke} S.,  {Dejonghe} H.,  {Zeilinger} W.~W.,   {Hau} G.~K.~T.,  2003,
  \mn@doi [\aap] {10.1051/0004-6361:20021866}, \href
  {http://adsabs.harvard.edu/abs/2003A%26A...400..119D} {400, 119}

\bibitem[\protect\citeauthoryear{{Faber}}{{Faber}}{1973}]{1973ApJ...179..423F}
{Faber} S.~M.,  1973, \mn@doi [\apj] {10.1086/151881}, \href
  {http://adsabs.harvard.edu/abs/1973ApJ...179..423F} {179, 423}

\bibitem[\protect\citeauthoryear{{Ferrarese} et~al.,}{{Ferrarese}
  et~al.}{2006}]{2006ApJS..164..334F}
{Ferrarese} L.,  et~al., 2006, \mn@doi [\apjs] {10.1086/501350}, \href
  {http://adsabs.harvard.edu/abs/2006ApJS..164..334F} {164, 334}

\bibitem[\protect\citeauthoryear{{Firth}, {Evstigneeva}, {Jones}, {Drinkwater},
  {Phillipps}  \& {Gregg}}{{Firth} et~al.}{2006}]{2006MNRAS.372.1856F}
{Firth} P.,  {Evstigneeva} E.~A.,  {Jones} J.~B.,  {Drinkwater} M.~J.,
  {Phillipps} S.,   {Gregg} M.~D.,  2006, \mn@doi [\mnras]
  {10.1111/j.1365-2966.2006.10993.x}, \href
  {http://adsabs.harvard.edu/abs/2006MNRAS.372.1856F} {372, 1856}

\bibitem[\protect\citeauthoryear{{Forbes}, {Lasky}, {Graham}  \&
  {Spitler}}{{Forbes} et~al.}{2008}]{2008MNRAS.389.1924F}
{Forbes} D.~A.,  {Lasky} P.,  {Graham} A.~W.,   {Spitler} L.,  2008, \mn@doi
  [\mnras] {10.1111/j.1365-2966.2008.13739.x}, \href
  {http://adsabs.harvard.edu/abs/2008MNRAS.389.1924F} {389, 1924}

\bibitem[\protect\citeauthoryear{{Forbes}, {Spitler}, {Graham}, {Foster}, {Hau}
   \& {Benson}}{{Forbes} et~al.}{2011}]{2011MNRAS.413.2665F}
{Forbes} D.~A.,  {Spitler} L.~R.,  {Graham} A.~W.,  {Foster} C.,  {Hau}
  G.~K.~T.,   {Benson} A.,  2011, \mn@doi [\mnras]
  {10.1111/j.1365-2966.2011.18335.x}, \href
  {http://adsabs.harvard.edu/abs/2011MNRAS.413.2665F} {413, 2665}

\bibitem[\protect\citeauthoryear{{Forbes}, {Pota}, {Usher}, {Strader},
  {Romanowsky}, {Brodie}, {Arnold}  \& {Spitler}}{{Forbes}
  et~al.}{2013}]{2013MNRAS.435L...6F}
{Forbes} D.~A.,  {Pota} V.,  {Usher} C.,  {Strader} J.,  {Romanowsky} A.~J.,
  {Brodie} J.~P.,  {Arnold} J.~A.,   {Spitler} L.~R.,  2013, \mn@doi [\mnras]
  {10.1093/mnrasl/slt078}, \href
  {http://adsabs.harvard.edu/abs/2013MNRAS.435L...6F} {435, L6}

\bibitem[\protect\citeauthoryear{{Forbes}, {Norris}, {Strader}, {Romanowsky},
  {Pota}, {Kannappan}, {Brodie}  \& {Huxor}}{{Forbes}
  et~al.}{2014}]{2014MNRAS.444.2993F}
{Forbes} D.~A.,  {Norris} M.~A.,  {Strader} J.,  {Romanowsky} A.~J.,  {Pota}
  V.,  {Kannappan} S.~J.,  {Brodie} J.~P.,   {Huxor} A.,  2014, \mn@doi
  [\mnras] {10.1093/mnras/stu1631}, \href
  {http://adsabs.harvard.edu/abs/2014MNRAS.444.2993F} {444, 2993}

\bibitem[\protect\citeauthoryear{{Geha}, {Guhathakurta}  \& {van der
  Marel}}{{Geha} et~al.}{2002}]{2002AJ....124.3073G}
{Geha} M.,  {Guhathakurta} P.,   {van der Marel} R.~P.,  2002, \mn@doi [\aj]
  {10.1086/344764}, \href {http://adsabs.harvard.edu/abs/2002AJ....124.3073G}
  {124, 3073}

\bibitem[\protect\citeauthoryear{{Geha}, {Guhathakurta}  \& {van der
  Marel}}{{Geha} et~al.}{2003}]{2003AJ....126.1794G}
{Geha} M.,  {Guhathakurta} P.,   {van der Marel} R.~P.,  2003, \mn@doi [\aj]
  {10.1086/377624}, \href {http://adsabs.harvard.edu/abs/2003AJ....126.1794G}
  {126, 1794}

\bibitem[\protect\citeauthoryear{{Gu{\'e}rou} et~al.,}{{Gu{\'e}rou}
  et~al.}{2015}]{2015ApJ...804...70G}
{Gu{\'e}rou} A.,  et~al., 2015, \mn@doi [\apj] {10.1088/0004-637X/804/1/70},
  \href {http://adsabs.harvard.edu/abs/2015ApJ...804...70G} {804, 70}

\bibitem[\protect\citeauthoryear{{Janz} \& {Lisker}}{{Janz} \&
  {Lisker}}{2008}]{2008ApJ...689L..25J}
{Janz} J.,  {Lisker} T.,  2008, \mn@doi [\apjl] {10.1086/595720}, \href
  {http://adsabs.harvard.edu/abs/2008ApJ...689L..25J} {689, L25}

\bibitem[\protect\citeauthoryear{{Janz} \& {Lisker}}{{Janz} \&
  {Lisker}}{2009}]{2009ApJ...696L.102J}
{Janz} J.,  {Lisker} T.,  2009, \mn@doi [\apjl] {10.1088/0004-637X/696/1/L102},
  \href {http://adsabs.harvard.edu/abs/2009ApJ...696L.102J} {696, L102}

\bibitem[\protect\citeauthoryear{{Jedrzejewski}}{{Jedrzejewski}}{1987}]{1987MNRAS.226..747J}
{Jedrzejewski} R.~I.,  1987, \mnras, \href
  {http://adsabs.harvard.edu/abs/1987MNRAS.226..747J} {226, 747}

\bibitem[\protect\citeauthoryear{{Jerjen}, {Binggeli}  \& {Barazza}}{{Jerjen}
  et~al.}{2004}]{2004AJ....127..771J}
{Jerjen} H.,  {Binggeli} B.,   {Barazza} F.~D.,  2004, \mn@doi [\aj]
  {10.1086/381065}, \href {http://adsabs.harvard.edu/abs/2004AJ....127..771J}
  {127, 771}

\bibitem[\protect\citeauthoryear{{Jorgensen}, {Franx}  \&
  {Kjaergaard}}{{Jorgensen} et~al.}{1995}]{1995MNRAS.276.1341J}
{Jorgensen} I.,  {Franx} M.,   {Kjaergaard} P.,  1995, \mnras, \href
  {http://adsabs.harvard.edu/abs/1995MNRAS.276.1341J} {276, 1341}

\bibitem[\protect\citeauthoryear{{Kim} et~al.,}{{Kim}
  et~al.}{2014}]{2014ApJS..215...22K}
{Kim} S.,  et~al., 2014, \mn@doi [\apjs] {10.1088/0067-0049/215/2/22}, \href
  {http://adsabs.harvard.edu/abs/2014ApJS..215...22K} {215, 22}

\bibitem[\protect\citeauthoryear{{Kormendy} \& {Bender}}{{Kormendy} \&
  {Bender}}{2012}]{2012ApJS..198....2K}
{Kormendy} J.,  {Bender} R.,  2012, \mn@doi [\apjs]
  {10.1088/0067-0049/198/1/2}, \href
  {http://adsabs.harvard.edu/abs/2012ApJS..198....2K} {198, 2}

\bibitem[\protect\citeauthoryear{{Kormendy}, {Fisher}, {Cornell}  \&
  {Bender}}{{Kormendy} et~al.}{2009}]{2009ApJS..182..216K}
{Kormendy} J.,  {Fisher} D.~B.,  {Cornell} M.~E.,   {Bender} R.,  2009, \mn@doi
  [\apjs] {10.1088/0067-0049/182/1/216}, \href
  {http://adsabs.harvard.edu/abs/2009ApJS..182..216K} {182, 216}

\bibitem[\protect\citeauthoryear{{Lisker}, {Grebel}  \& {Binggeli}}{{Lisker}
  et~al.}{2008}]{2008AJ....135..380L}
{Lisker} T.,  {Grebel} E.~K.,   {Binggeli} B.,  2008, \mn@doi [\aj]
  {10.1088/0004-6256/135/1/380}, \href
  {http://adsabs.harvard.edu/abs/2008AJ....135..380L} {135, 380}

\bibitem[\protect\citeauthoryear{{Maraston}}{{Maraston}}{2005}]{2005MNRAS.362..799M}
{Maraston} C.,  2005, \mn@doi [\mnras] {10.1111/j.1365-2966.2005.09270.x},
  \href {http://adsabs.harvard.edu/abs/2005MNRAS.362..799M} {362, 799}

\bibitem[\protect\citeauthoryear{{Mastropietro}, {Moore}, {Mayer},
  {Debattista}, {Piffaretti}  \& {Stadel}}{{Mastropietro}
  et~al.}{2005}]{2005MNRAS.364..607M}
{Mastropietro} C.,  {Moore} B.,  {Mayer} L.,  {Debattista} V.~P.,  {Piffaretti}
  R.,   {Stadel} J.,  2005, \mn@doi [\mnras]
  {10.1111/j.1365-2966.2005.09579.x}, \href
  {http://adsabs.harvard.edu/abs/2005MNRAS.364..607M} {364, 607}

\bibitem[\protect\citeauthoryear{{Mayer}, {Governato}, {Colpi}, {Moore},
  {Quinn}  \& {Baugh}}{{Mayer} et~al.}{2001a}]{2001Ap&SS.276..375M}
{Mayer} L.,  {Governato} F.,  {Colpi} M.,  {Moore} B.,  {Quinn} T.~R.,
  {Baugh} C.~M.,  2001a, \mn@doi [\apss] {10.1023/A:1017562225479}, \href
  {http://adsabs.harvard.edu/abs/2001Ap%26SS.276..375M} {276, 375}

\bibitem[\protect\citeauthoryear{{Mayer}, {Governato}, {Colpi}, {Moore},
  {Quinn}, {Wadsley}, {Stadel}  \& {Lake}}{{Mayer}
  et~al.}{2001b}]{2001ApJ...547L.123M}
{Mayer} L.,  {Governato} F.,  {Colpi} M.,  {Moore} B.,  {Quinn} T.,  {Wadsley}
  J.,  {Stadel} J.,   {Lake} G.,  2001b, \mn@doi [\apjl] {10.1086/318898},
  \href {http://adsabs.harvard.edu/abs/2001ApJ...547L.123M} {547, L123}

\bibitem[\protect\citeauthoryear{{Mayer}, {Mastropietro}, {Wadsley}, {Stadel}
  \& {Moore}}{{Mayer} et~al.}{2006}]{2006MNRAS.369.1021M}
{Mayer} L.,  {Mastropietro} C.,  {Wadsley} J.,  {Stadel} J.,   {Moore} B.,
  2006, \mn@doi [\mnras] {10.1111/j.1365-2966.2006.10403.x}, \href
  {http://adsabs.harvard.edu/abs/2006MNRAS.369.1021M} {369, 1021}

\bibitem[\protect\citeauthoryear{{McConnachie}}{{McConnachie}}{2012}]{2012AJ....144....4M}
{McConnachie} A.~W.,  2012, \mn@doi [\aj] {10.1088/0004-6256/144/1/4}, \href
  {http://adsabs.harvard.edu/abs/2012AJ....144....4M} {144, 4}

\bibitem[\protect\citeauthoryear{{Misgeld} \& {Hilker}}{{Misgeld} \&
  {Hilker}}{2011}]{2011MNRAS.414.3699M}
{Misgeld} I.,  {Hilker} M.,  2011, \mn@doi [\mnras]
  {10.1111/j.1365-2966.2011.18669.x}, \href
  {http://adsabs.harvard.edu/abs/2011MNRAS.414.3699M} {414, 3699}

\bibitem[\protect\citeauthoryear{{Moore}, {Katz}, {Lake}, {Dressler}  \&
  {Oemler}}{{Moore} et~al.}{1996}]{1996Natur.379..613M}
{Moore} B.,  {Katz} N.,  {Lake} G.,  {Dressler} A.,   {Oemler} A.,  1996,
  \mn@doi [\nat] {10.1038/379613a0}, \href
  {http://adsabs.harvard.edu/abs/1996Natur.379..613M} {379, 613}

\bibitem[\protect\citeauthoryear{{Norris} et~al.,}{{Norris}
  et~al.}{2014}]{2014MNRAS.443.1151N}
{Norris} M.~A.,  et~al., 2014, \mn@doi [\mnras] {10.1093/mnras/stu1186}, \href
  {http://adsabs.harvard.edu/abs/2014MNRAS.443.1151N} {443, 1151}

\bibitem[\protect\citeauthoryear{{Peng} et~al.,}{{Peng}
  et~al.}{2008}]{2008ApJ...681..197P}
{Peng} E.~W.,  et~al., 2008, \mn@doi [\apj] {10.1086/587951}, \href
  {http://adsabs.harvard.edu/abs/2008ApJ...681..197P} {681, 197}

\bibitem[\protect\citeauthoryear{{Penny}, {Conselice}, {de Rijcke}  \&
  {Held}}{{Penny} et~al.}{2009}]{2009MNRAS.393.1054P}
{Penny} S.~J.,  {Conselice} C.~J.,  {de Rijcke} S.,   {Held} E.~V.,  2009,
  \mn@doi [\mnras] {10.1111/j.1365-2966.2008.14269.x}, \href
  {http://adsabs.harvard.edu/abs/2009MNRAS.393.1054P} {393, 1054}

\bibitem[\protect\citeauthoryear{{Penny}, {Forbes}, {Strader}, {Usher},
  {Brodie}  \& {Romanowsky}}{{Penny} et~al.}{2014a}]{2014MNRAS.439.3808P}
{Penny} S.~J.,  {Forbes} D.~A.,  {Strader} J.,  {Usher} C.,  {Brodie} J.~P.,
  {Romanowsky} A.~J.,  2014a, \mn@doi [\mnras] {10.1093/mnras/stu232}, \href
  {http://adsabs.harvard.edu/abs/2014MNRAS.439.3808P} {439, 3808}

\bibitem[\protect\citeauthoryear{{Penny}, {Forbes}, {Pimbblet}  \&
  {Floyd}}{{Penny} et~al.}{2014b}]{2014MNRAS.443.3381P}
{Penny} S.~J.,  {Forbes} D.~A.,  {Pimbblet} K.~A.,   {Floyd} D.~J.~E.,  2014b,
  \mn@doi [\mnras] {10.1093/mnras/stu1397}, \href
  {http://adsabs.harvard.edu/abs/2014MNRAS.443.3381P} {443, 3381}

\bibitem[\protect\citeauthoryear{{Pfeffer} \& {Baumgardt}}{{Pfeffer} \&
  {Baumgardt}}{2013}]{2013MNRAS.433.1997P}
{Pfeffer} J.,  {Baumgardt} H.,  2013, \mn@doi [\mnras] {10.1093/mnras/stt867},
  \href {http://adsabs.harvard.edu/abs/2013MNRAS.433.1997P} {433, 1997}

\bibitem[\protect\citeauthoryear{{Pointecouteau}, {Arnaud}  \&
  {Pratt}}{{Pointecouteau} et~al.}{2005}]{2005A&A...435....1P}
{Pointecouteau} E.,  {Arnaud} M.,   {Pratt} G.~W.,  2005, \mn@doi [\aap]
  {10.1051/0004-6361:20042569}, \href
  {http://adsabs.harvard.edu/abs/2005A%26A...435....1P} {435, 1}

\bibitem[\protect\citeauthoryear{{Ry{\'s}}, {Falc{\'o}n-Barroso}  \& {van de
  Ven}}{{Ry{\'s}} et~al.}{2013}]{2013MNRAS.428.2980R}
{Ry{\'s}} A.,  {Falc{\'o}n-Barroso} J.,   {van de Ven} G.,  2013, \mn@doi
  [\mnras] {10.1093/mnras/sts245}, \href
  {http://adsabs.harvard.edu/abs/2013MNRAS.428.2980R} {428, 2980}

\bibitem[\protect\citeauthoryear{{Ry{\'s}}, {van de Ven}  \&
  {Falc{\'o}n-Barroso}}{{Ry{\'s}} et~al.}{2014}]{2014MNRAS.439..284R}
{Ry{\'s}} A.,  {van de Ven} G.,   {Falc{\'o}n-Barroso} J.,  2014, \mn@doi
  [\mnras] {10.1093/mnras/stt2417}, \href
  {http://adsabs.harvard.edu/abs/2014MNRAS.439..284R} {439, 284}

\bibitem[\protect\citeauthoryear{{Sheinis}, {Bolte}, {Epps}, {Kibrick},
  {Miller}, {Radovan}, {Bigelow}  \& {Sutin}}{{Sheinis}
  et~al.}{2002}]{2002PASP..114..851S}
{Sheinis} A.~I.,  {Bolte} M.,  {Epps} H.~W.,  {Kibrick} R.~I.,  {Miller} J.~S.,
   {Radovan} M.~V.,  {Bigelow} B.~C.,   {Sutin} B.~M.,  2002, \mn@doi [\pasp]
  {10.1086/341706}, \href {http://adsabs.harvard.edu/abs/2002PASP..114..851S}
  {114, 851}

\bibitem[\protect\citeauthoryear{{Smith Castelli}, {Gonz{\'a}lez}, {Faifer}  \&
  {Forte}}{{Smith Castelli} et~al.}{2013}]{2013ApJ...772...68S}
{Smith Castelli} A.~V.,  {Gonz{\'a}lez} N.~M.,  {Faifer} F.~R.,   {Forte}
  J.~C.,  2013, \mn@doi [\apj] {10.1088/0004-637X/772/1/68}, \href
  {http://adsabs.harvard.edu/abs/2013ApJ...772...68S} {772, 68}

\bibitem[\protect\citeauthoryear{{Smith} et~al.,}{{Smith}
  et~al.}{2002}]{2002AJ....123.2121S}
{Smith} J.~A.,  et~al., 2002, \mn@doi [\aj] {10.1086/339311}, \href
  {http://adsabs.harvard.edu/abs/2002AJ....123.2121S} {123, 2121}

\bibitem[\protect\citeauthoryear{{Smith}, {Lucey}, {Price}, {Hudson}  \&
  {Phillipps}}{{Smith} et~al.}{2012}]{2012MNRAS.419.3167S}
{Smith} R.~J.,  {Lucey} J.~R.,  {Price} J.,  {Hudson} M.~J.,   {Phillipps} S.,
  2012, \mn@doi [\mnras] {10.1111/j.1365-2966.2011.19956.x}, \href
  {http://adsabs.harvard.edu/abs/2012MNRAS.419.3167S} {419, 3167}

\bibitem[\protect\citeauthoryear{{Tollerud}, {Bullock}, {Graves}  \&
  {Wolf}}{{Tollerud} et~al.}{2011}]{2011ApJ...726..108T}
{Tollerud} E.~J.,  {Bullock} J.~S.,  {Graves} G.~J.,   {Wolf} J.,  2011,
  \mn@doi [\apj] {10.1088/0004-637X/726/2/108}, \href
  {http://adsabs.harvard.edu/abs/2011ApJ...726..108T} {726, 108}

\bibitem[\protect\citeauthoryear{{Toloba}, {Boselli}, {Cenarro}, {Peletier},
  {Gorgas}, {Gil de Paz}  \& {Mu{\~n}oz-Mateos}}{{Toloba}
  et~al.}{2011}]{2011A&A...526A.114T}
{Toloba} E.,  {Boselli} A.,  {Cenarro} A.~J.,  {Peletier} R.~F.,  {Gorgas} J.,
  {Gil de Paz} A.,   {Mu{\~n}oz-Mateos} J.~C.,  2011, \mn@doi [\aap]
  {10.1051/0004-6361/201015344}, \href
  {http://adsabs.harvard.edu/abs/2011A%26A...526A.114T} {526, A114}

\bibitem[\protect\citeauthoryear{{Toloba}, {Boselli}, {Peletier},
  {Falc{\'o}n-Barroso}, {van de Ven}  \& {Gorgas}}{{Toloba}
  et~al.}{2012}]{2012A&A...548A..78T}
{Toloba} E.,  {Boselli} A.,  {Peletier} R.~F.,  {Falc{\'o}n-Barroso} J.,  {van
  de Ven} G.,   {Gorgas} J.,  2012, \mn@doi [\aap]
  {10.1051/0004-6361/201218944}, \href
  {http://adsabs.harvard.edu/abs/2012A%26A...548A..78T} {548, A78}

\bibitem[\protect\citeauthoryear{{Toloba} et~al.,}{{Toloba}
  et~al.}{2014}]{2014ApJS..215...17T}
{Toloba} E.,  et~al., 2014, \mn@doi [\apjs] {10.1088/0067-0049/215/2/17}, \href
  {http://adsabs.harvard.edu/abs/2014ApJS..215...17T} {215, 17}

\bibitem[\protect\citeauthoryear{{Toloba} et~al.,}{{Toloba}
  et~al.}{2015}]{2014arXiv1410.1552T}
{Toloba} E.,  et~al., 2015, \mn@doi [\apj] {10.1088/0004-637X/799/2/172}, \href
  {http://adsabs.harvard.edu/abs/2015ApJ...799..172T} {799, 172}

\bibitem[\protect\citeauthoryear{{Trentham} \& {Tully}}{{Trentham} \&
  {Tully}}{2009}]{2009MNRAS.398..722T}
{Trentham} N.,  {Tully} R.~B.,  2009, \mn@doi [\mnras]
  {10.1111/j.1365-2966.2009.15189.x}, \href
  {http://adsabs.harvard.edu/abs/2009MNRAS.398..722T} {398, 722}

\bibitem[\protect\citeauthoryear{{Trentham}, {Tully}  \& {Mahdavi}}{{Trentham}
  et~al.}{2006}]{2006MNRAS.369.1375T}
{Trentham} N.,  {Tully} R.~B.,   {Mahdavi} A.,  2006, \mn@doi [\mnras]
  {10.1111/j.1365-2966.2006.10378.x}, \href
  {http://adsabs.harvard.edu/abs/2006MNRAS.369.1375T} {369, 1375}

\bibitem[\protect\citeauthoryear{{Wolf}, {Martinez}, {Bullock}, {Kaplinghat},
  {Geha}, {Mu{\~n}oz}, {Simon}  \& {Avedo}}{{Wolf}
  et~al.}{2010}]{2010MNRAS.406.1220W}
{Wolf} J.,  {Martinez} G.~D.,  {Bullock} J.~S.,  {Kaplinghat} M.,  {Geha} M.,
  {Mu{\~n}oz} R.~R.,  {Simon} J.~D.,   {Avedo} F.~F.,  2010, \mn@doi [\mnras]
  {10.1111/j.1365-2966.2010.16753.x}, \href
  {http://adsabs.harvard.edu/abs/2010MNRAS.406.1220W} {406, 1220}

\bibitem[\protect\citeauthoryear{{York} et~al.,}{{York}
  et~al.}{2000}]{2000AJ....120.1579Y}
{York} D.~G.,  et~al., 2000, \mn@doi [\aj] {10.1086/301513}, \href
  {http://adsabs.harvard.edu/abs/2000AJ....120.1579Y} {120, 1579}

\makeatother
\end{thebibliography}





\bsp	
\label{lastpage}
\end{document}